\documentclass[12pt]{article}
\pdfoutput=1

\newcommand{\VersionInformation}{}  
\InputIfFileExists{Debug}{}{}

\usepackage{jheppub}
\usepackage[utf8]{inputenc}
\usepackage{amsmath}
\usepackage{amsthm}
\usepackage{amsfonts}          
\usepackage{amssymb}           
\usepackage[mathscr]{eucal}
\usepackage{graphicx}
\usepackage{color}
\usepackage{hypbmsec}
\usepackage{fancyvrb}
\usepackage{sepnum}
\usepackage{xspace}
\usepackage{booktabs}
\usepackage{rotating}
\usepackage{multirow}
\usepackage{verbatim}
\usepackage[all]{xy}
\usepackage[isu,bf]{caption}
\usepackage{array}
\usepackage{url}

\newlength{\xtrawidth}
\newlength{\xtraheight}
\ifx\DEBUG\EMPTYMACRO
\else
  \usepackage{showlabels}
\fi

\def\clap#1{\hbox to 0pt{\hss#1\hss}}

\makeatletter
  \def\adots{\mathinner{\mkern2mu\raise\p@\hbox{.}
      \mkern2mu\raise4\p@\hbox{.}\mkern1mu
      \raise7\p@\vbox{\kern7\p@\hbox{.}}\mkern1mu}}
\makeatother

\newcommand{\eqdef}{%
  \mathrel{\lower.1mm
    \hbox{$\stackrel{\lower.424ex\hbox{\scriptsize def}}{=}$}}
}

\newcommand{\C}{\ensuremath{{\mathbb{C}}}}

\newcommand{\Z}{\mathbb{Z}}
\newcommand{\CP}{{\ensuremath{\mathop{\null {\mathbb{P}}}\nolimits}}}

\newcommand{\ptset}{\ensuremath{\{\text{pt.}\}}}

\DeclareMathOperator{\MW}{MW}

\DeclareMathOperator{\Span}{span}

\DeclareMathOperator{\Sym}{Sym}

\DeclareMathOperator{\img}{img}

\DeclareMathOperator{\Res}{Res}

\newcommand{\rep}[1]{\ensuremath{\mathbf{#1}}}

\DeclareMathOperator{\rk}{rk}

\newcommand{\Lsheaf}{\ensuremath{\mathscr{L}}}
\newcommand{\Rsheaf}{\ensuremath{\mathscr{R}}}
\newcommand{\Osheaf}{\ensuremath{\mathscr{O}}}

\newenvironment{descriptionlist}{%
\begin{list}%
{}%
{}}%
{\end{list}}

{\everymath{\displaystyle\everymath{}}\array}%
{\endarray}

\usepackage{color}
\usepackage{listings} 
\usepackage{courier}

\definecolor{dbluecolor}{rgb}{0.01,0.02,0.7}
\definecolor{dgreencolor}{rgb}{0.2,0.4,0.0}
\definecolor{dgraycolor}{rgb}{0.30,0.3,0.30}

\lstdefinelanguage{SageInputLanguage}{
  language=Python, 
  morekeywords={False,sage,True},
  sensitive=true,
}

\lstdefinestyle{SageInput}{
  language=SageInputLanguage,
  basicstyle=\fontsize{11pt}{11pt}\ttfamily\bfseries,
  commentstyle={\ttfamily\color{dgreencolor}},
  stringstyle={\color{dgraycolor}\bfseries},
  keywordstyle=\ttfamily\color{dbluecolor}\bfseries\color{red},
  xleftmargin=25pt,
  belowskip=3pt,
}

\lstdefinelanguage{SageOutputLanguage}{
  morekeywords={False,True},
  sensitive=true,
}

\lstdefinestyle{SageOutput}{
  language=SageOutputLanguage,
  basicstyle={\fontsize{11pt}{11pt}\ttfamily},
  commentstyle={\ttfamily\color{dgreencolor}},
  keywordstyle={\ttfamily\color{dbluecolor}},
  stringstyle={\ttfamily\color{dgraycolor}},
  xleftmargin=25pt,
  aboveskip=0pt,
}

\lstdefinestyle{DefaultSageInputOutput}{
  identifierstyle=,
  numbersep=5pt,
  aboveskip=0pt,
  belowskip=0pt,
  breaklines=true,
  numberstyle=\footnotesize,
  numbers=right,
}

\begin{document}

\begin{titlepage}
\begin{flushright}
\parbox[t]{1.8in}{\begin{flushright} MPP-2014-548 \end{flushright}}
\end{flushright}

\begin{center}

\vspace*{ 1.2cm}

\textbf{\LARGE\boldmath 
  Complete Intersection Fibers in F-Theory
}

\vskip 1cm

\renewcommand{\thefootnote}{}
\begin{center}
  Volker Braun,${}^1$ 
  Thomas W.~Grimm,${}^2$
  and Jan Keitel${}^2$ 
  \footnote{\texttt{volker.braun@maths.ox.ac.uk}; \texttt{grimm}, \texttt{jkeitel@mpp.mpg.de}}
\end{center}
\vskip .2cm
\renewcommand{\thefootnote}{\arabic{footnote}}

{${}^1$Mathematical Institute,
  University of Oxford\\
  24-29 St Giles', Oxford, OX1 3LB, United Kingdom}

\vspace*{.2cm}
 
{${}^2$Max-Planck-Institut f\"ur Physik, \\
F\"ohringer Ring 6, 80805 Munich, Germany}

 \vspace*{.8cm}

\end{center}

\vskip 0.2cm
 
\begin{center} {\bf ABSTRACT } \end{center}

Global F-theory compactifications whose fibers are realized as
complete intersections form a richer set of models than just
hypersurfaces. The detailed study of the physics associated with such
geometries depends crucially on being able to put the elliptic fiber
into Weierstrass form.  While such a transformation is always
guaranteed to exist, its explicit form is only known in a few special
cases. We present a general algorithm for computing the Weierstrass
form of elliptic curves defined as complete intersections of different
codimensions and use it to solve all cases of complete intersections
of two equations in an ambient toric variety. Using this result, we
determine the toric Mordell-Weil groups of all $3134$ nef partitions
obtained from the $4319$ three-dimensional reflexive polytopes and
find new groups that do not exist for toric hypersurfaces.  As an
application, we construct several models that cannot be realized as
toric hypersurfaces, such as the first toric $SU(5)$ GUT model in the
literature with distinctly charged $\rep{10}$ representations and an
F-theory model with discrete gauge group $\mathbb{Z}_4$ whose dual
fiber has a Mordell-Weil group with $\mathbb{Z}_4$ torsion.

\vspace*{.5cm}

\hfill {November 12, 2014}
\end{titlepage}
\setcounter{footnote}{0}
\VersionInformation

\tableofcontents


 \newpage
 
\section{Introduction}
\label{sec:intro}

F-theory \cite{Vafa:1996xn} provides a convenient way of realizing the $SL(2,\mathbb{Z})$ symmetry
of Type IIB string theory geometrically by relating it to the modular group
acting on the complex structure of a $T^2$. In particular,
the complex structure of this auxiliary two-torus is identified with the axio-dilaton
of the low-energy effective action.
For Calabi-Yau manifolds
that are non-trivial $T^2$-fibrations one thus obtains a geometric
description of a Type IIB background with varying axio-dilaton $\tau$.

Since the axio-dilaton diverges at the position of D7-branes in the Type IIB compactification,
$\tau$ contains information about the low-energy effective theory and is therefore
one of the main quantities of interest. $\tau$ and especially the locus of its singularities
can easily be obtained if the defining equation of the $T^2$ is given in Weierstrass form
\begin{align} \label{e:wf}
 y^2 = x^3 + f x + g\,.
\end{align}
In general, for every torus fibration with a global section a map into
this form is guaranteed to exist. If the fibration
does not have a global section, then one can replace the genus-one
curve by its Jacobian, which is then guaranteed to have a section while
maintaining the same discriminant.\footnote{However, the Jacobian might have terminal
singularities even if the original fibration was smooth~\cite{Braun:2014oya}.}
In practice, however, finding this map can be challenging
and the solution to this problem is only known in a few special cases. The simplest of these
cases is the elliptic curve inside $\mathbb{P}_{231}$ whose generic form is given by
\begin{align} \label{e:p123}
 y^2 + a_1 x y z + a_3 y z^3 = x^3 + a_2 x^2 z^2 + a_4 x z^6 + a_6 z^6\,.
\end{align}
Equation \eqref{e:p123} can be brought into Weierstrass form simply by completing the square
and the cube with respect to $y^2$ and $x^3$. Possibly for this reason,
much of the early F-theory literature focused on such scenarios and constructed Calabi-Yau
manifolds inside $\mathbb{P}_{231}$ fibrations over $B'$, with the $T^2$ a hypersurface in $\mathbb{P}_{231}$
and the base $B_{n-1}$ a complete intersection in $B'$.
In order to harness the full power of algebraic geometry, one ordinarily considers complete intersections
whose defining equations have generic coefficients inside such a space. As soon as one does so,
however, considering only fibers embedded in $\mathbb{P}_{231}$ heavily restricts the
low-energy effective physics of the corresponding F-theory compactifications. In particular,
generic fibers inside $\mathbb{P}_{231}$ do not lead to Abelian gauge factors.
In recent years, the original focus on engineering non-Abelian gauge theories in global F-theory
\cite{Candelas:1996su,Bershadsky:1996nh,Candelas:1997eh}
has shifted towards advancing the understanding of their Abelian counterparts.
As a consequence, it has become necessary to consider more general fiber embeddings, starting with
a blow-up of $\mathbb{P}_{231}$ in \cite{Grimm:2010ez}, extended to more general cases with a single $U(1)$ in
\cite{Morrison:2012ei,Braun:2013yti,Grimm:2013oga,Kuntzler:2014ila} and finally progressing to higher-rank $U(1)$s
\cite{Borchmann:2013hta,Cvetic:2013nia,Cvetic:2013uta,Borchmann:2013jwa,Cvetic:2013qsa}
and a treatment of embeddings in all $16$ toric surfaces in \cite{Braun:2013nqa,Klevers:2014bqa}.
Most recently, torus fibers that do not generically have a section, i.e.~genus-one curves
that are not elliptic curves, have started to be investigated in
\cite{Braun:2014oya,Morrison:2014era,Anderson:2014yva,Klevers:2014bqa,Garcia-Etxebarria:2014qua,
Mayrhofer:2014haa, Mayrhofer:2014laa}. Furthermore, progress has been made in
also understanding geometrically massive $U(1)$s \cite{Grimm:2011tb,Braun:2014nva}.

With the exception of \cite{Cvetic:2013qsa},
in which purely Abelian $U(1)^3$ models were studied, and \cite{Mayrhofer:2012zy} where an $SU(5)$ singularity
was resolved using a complete intersection, all of these works have embedded the elliptic fiber as a \emph{hypersurface}
in a two-dimensional toric variety.
For these cases, computing the Weierstrass form was developed
in \cite{Braun:2011ux}. However, as shown in \cite{Braun:2013nqa}, this still imposes a considerable
constraint on the resulting F-theory models. Apart from limiting the toric Mordell-Weil group to rank $\leq 3$,
the fact that the elliptic curve is a hypersurface in an ambient variety also restricts
the possible resolutions of non-Abelian singularities. In particular, with respect to $SU(5)$ GUTs,
it implies that there exists only a single antisymmetric matter representation in the spectrum
of the low-energy effective theory.
The restriction on the matter content applies of course only to \emph{resolved} manifolds --- after blowing down, singular
models can be constructed as hypersurfaces, as is obvious from the fact that there exists a transformation
to Weierstrass form. The singularity enhancements of Calabi-Yau manifolds with two sections
were studied systematically in \cite{Kuntzler:2014ila}.

In this work, we aim to extend the effort of \cite{Braun:2011ux} and provide a new method for
bringing a large class of complete intersection fibers into Weierstrass form. This class contains
both models without section and with section(s).
As alluded to above,
we compute the Weierstrass form of the associated Jacobian in the cases which do not have a section.
We develop the algorithm in \autoref{s:algorithm} after giving a short
summary of some of the mathematical background in \autoref{s:koszul}. In \autoref{s:classification} we then review
complete intersections in toric varieties and, as an application of our algorithm,
classify all toric Mordell-Weil groups of the $3134$ nef partitions of the $4319$ three-dimensional
reflexive polytopes.
Since the full list of results is too long to be included in the text of this paper, we have created a website at
\begin{align}
  \textrm{\url{http://wwwth.mpp.mpg.de/members/jkeitel/Weierstrass/}}
\end{align}
with a database of the $3134$ nef partitions of three-dimensional reflexive polyhedra,
their Weierstrass forms, toric Mordell-Weil groups and generic non-Abelian singularities.
Finally, in \autoref{s:examples} we showcase several example manifolds that exhibit
features not present for elliptic fibers that are hypersurface. Among these are a manifold
with Mordell-Weil torsion $\mathbb{Z}_4$ and an F-theory model with discrete gauge group $\mathbb{Z}_4$.
Furthermore we demonstrate that considering complete intersection fibers
indeed evades the no-go theorem of \cite{Braun:2013nqa} and present the first torically realized
$SU(5) \times U(1)^2$ model with distinctly charged antisymmetric matter representations.


 \section{Koszul and Residues} \label{s:koszul}

The one indispensable tool for studying complete intersections is
the Koszul complex and the associated hypercohomology spectral
sequence. In the interest of a self-contained presentation let us
quickly review these. Of course we have nothing new to say about
these~\cite{Berglund:1994my}, the cognoscenti are advised to skip to
\autoref{s:algorithm}.

The simplest way to think of line bundle valued cohomology groups
$H^k\big(\CP^d,\Osheaf(n)\big)$ is as holomorphic degree-$k$
differential forms that transform like degree-$n$ homogeneous
polynomials under rescalings of the homogeneous coordinates. More
generally, we can consider multiple homogeneous rescalings which just
amounts to a toric variety $X$ and line bundle $\Lsheaf$. Then
$H^k(X,\Lsheaf)$ are holomorphic degree-$k$ differential forms,
transforming like homogeneous polynomials whose degree of homogeneity
determined by the line bundle $\Lsheaf$. Ultimately we are interested
in a Calabi-Yau submanifold $Y\subset X$ cut out by two\footnote{The
  whole discussion of this section generalizes to arbitrary
  codimension, but for simplicity we restrict ourselves to 
  codimension two.}
transverse polynomials $p_1=p_2=0$. There are three ways to obtain a
degree-$k$ differential form on $Y$:
\begin{enumerate}
\item Restriction of a degree-$k$ form on $X$,
\item Residue integration of a degree-$(k+1)$ form around a small
  circle around either $p_1=0$ or $p_2=0$, and 
\item Two-fold residue integration around $p_1=p_2=0$ of a
  degree-$(k+2)$ form.
\end{enumerate}
It is convenient to define the residue operators
\begin{math}
  \Res_j(\omega) = \frac{1}{2\pi i}\oint \frac{(p_j\omega)}{p_j} 
\end{math}
and split the potential contributions $E_1^{p,q}$ to
$H^{p+q}(Y,\Lsheaf|_Y)$ into $(-p)$-fold residues of $q$-forms. Note the
minus sign in the definition of $p$, as the residue operator has
differential degree $-1$. We also have to be careful with the degree
under homogeneous rescalings, as the residue operator $\Res_j$ has us
multiply by the homogeneous polynomial $p_j$. The polynomial $p_j$
defines a divisor $D_j=V(p_j) = \{p_j = 0\}$, and the cohomology
groups of the line bundle $\Osheaf(D_j)$ precisely involve
differential forms of the same degree of homogeneity as $p_j$. Hence,
the residue operator actually maps
\begin{equation}
  \Res_j:~
  H^{k+1}\big(X, \Lsheaf(-D_j)\big)
  \longrightarrow
  H^k\big(Y, \Lsheaf|_Y\big)
\end{equation}
Putting everything together, the potential contributions to the
cohomology for a $3$-dimensional toric variety $X$ fill out the
tableau
\begin{multline}
  E_1^{p, q}(\Lsheaf) = \\
  \vcenter{ 
    \def\wA{45mm} 
    \def\wB{68mm} 
    \def\wC{20mm} 
    \def\h{8mm} 
    \xymatrix@C=0mm@R=0mm{
      {\scriptstyle q=3} & 
      *=<\wA,\h>[F]{ H^3\big(X,\Lsheaf(-D_1-D_2)\big) } &
      *=<\wB,\h>[F]{ H^3\big(X,\Lsheaf(-D_1)\big) \oplus 
        H^3\big(X,\Lsheaf(-D_2\big) } & 
      *=<\wC,\h>[F]{ H^3(X,\Lsheaf) } 
      \\ {\scriptstyle q=2} & 
      *=<\wA,\h>[F]{ H^2\big(X,\Lsheaf(-D_1-D_2)\big) } &
      *=<\wB,\h>[F]{ H^2\big(X,\Lsheaf(-D_1)\big) \oplus 
        H^2\big(X,\Lsheaf(-D_2\big) } & 
      *=<\wC,\h>[F]{ H^2(X,\Lsheaf) } 
      \\ {\scriptstyle q=1} & 
      *=<\wA,\h>[F]{ H^1\big(X,\Lsheaf(-D_1-D_2)\big) } &
      *=<\wB,\h>[F]{ H^1\big(X,\Lsheaf(-D_1)\big) \oplus 
        H^1\big(X,\Lsheaf(-D_2\big) } & 
      *=<\wC,\h>[F]{ H^1(X,\Lsheaf) } 
      \\ {\scriptstyle q=0} & 
      *=<\wA,\h>[F]{ H^0\big(X,\Lsheaf(-D_1-D_2)\big) } & 
      *=<\wB,\h>[F]{ H^0\big(X,\Lsheaf(-D_1)\big) \oplus 
        H^0\big(X,\Lsheaf(-D_2\big) } & 
      *=<\wC,\h>[F]{ H^0\big(X,\Lsheaf\big) } 
      \\ & {\scriptstyle p=-2} & {\scriptstyle p=-1} & 
      {\scriptstyle p=0} 
    }} 
  \\
  \Rightarrow 
  H^{p+q}(Y, \Lsheaf|_Y).
\end{multline}
with the map to $H^{p+q}$ being either $\Res_1\Res_2$, $\Res_1\oplus
\Res_2$, or restriction for the three respective columns. That way,
the entries along the diagonal can contribute to
$H^{p+q}(Y,\Lsheaf|_Y)$, but we have no reason to believe that these
are all independent.

In particular, the restrictions of two different $k$-forms $\alpha_1$,
$\alpha_2$ may very well be cohomologous on $Y$, even if they are not
on $X$. Clearly, this is the case when $\alpha_1-\alpha_2 = d
\Res(\omega)$ for some $k$-form $\omega$. Similarly, two forms on $Y$
that came from different residues might be related by a double
residue. This is implemented by a nilpotent\footnote{That $d_1^2=0$
  requires a suitable sign choice; Schematically $d_1^{p=-2}=(p_1,
  p_2)$ and $d_1^{p=-1}=\left(
    \begin{smallmatrix}
      -p_2 \\ p_1
    \end{smallmatrix}\right)$.} differential $d_1:E_1^{p,q} \to E_1^{p+1,
  q}$. Only the cohomology with respect to $d_1$ has a chance of
contributing to $H^{p+q}(Y,\Lsheaf|_Y)$. We arrange the
$d_1$-cohomology groups in the $E_2$-tableau
\begin{equation}
  E_2^{p,q} = 
  \frac{
    \ker\big( d_1:E_1^{p,q} \to E_1^{p+1,q} \big)
  }{
    \img\big( d_1:E_1^{p-1,q} \to E_1^{p,q} \big)
  }.
\end{equation}
Unfortunately, this is not the end of it and even a $d_1$-cohomology
class need not survive to a non-zero element of
$H^{p+q}(Y,\Lsheaf|_Y)$. This is the case when two different $k$-forms
$\alpha_1$, $\alpha_2$ on $X$ are related via a double residue of a
$(k+1)$-form, $\alpha_1-\alpha_2 = d \Res_1\Res_2(\omega)$. This is
implemented by yet another nilpotent differential $d_2:E_2^{p,q} \to
E_2^{p+2, q-1}$. Its cohomology forms the entries of the $E_3$-tableau.

In general, a spectral sequence is an infinite sequence of tableaux
$E_i^{p,q}$ and differentials $d_i:E_i^{p,q} \to E_i^{p+i, q+1-i}$. In
the case of a two-fold complete intersection, this process stabilizes
at $E_3=E_\infty$ because all higher differentials are starting or
ending outside of the $3\times 4$ region with the non-zero
entries. The diagonals of the $E_\infty$ tableau are a filtration of
the cohomology groups $H^{p+q}(Y,\Lsheaf|_Y)$. In particular, this
implies that
\begin{equation}
  \dim H^k(Y,\Lsheaf|_Y) = \sum_{p+q = k} \dim E^{p,q}_\infty
\end{equation}
and therefore one can reconstruct the dimension of the line bundle
cohomology groups on the complete intersection from the knowledge
of the dimensions of the $E_\infty$ tableau entries.

\section{Weierstrass Form for Complete Intersections} 
\label{s:algorithm}

In this section, we develop an algorithm to bring an elliptic curve
defined by a complete intersection into Weierstrass form. The underlying
idea is spelled out in \autoref{ss:basic_algorithm}. In
\autoref{ss:line_bundles} and \autoref{sec:d2} we discuss the
relations between the line bundles on the complete intersection
and the line bundles on the ambient space.
Using an explicit example, we show in \autoref{ss:algorithm} explicitly
how to apply our algorithm in practice.
Finally, in \autoref{sec:except} we manually compute the Weierstrass
forms for the only two codimension two examples to which
the algorithm cannot be applied.

\subsection{Basic Algorithm} \label{ss:basic_algorithm}

We are interested in finding the Weierstrass form of an elliptic curve
over a base field that is not necessarily algebraically closed. In
particular, if the base field is the function field of the base then
this includes the case of elliptic fibrations. There are two different
ways of quantifying how complicated the ambient space is: One is going
from hypersurfaces to complete intersections to general subvarieties
whose number of defining equations exceeds their codimension. This is
convenient for constructing smooth Calabi-Yau manifolds, since we can
often use genericity of the defining equations to argue that a generic
subvariety is smooth. As far as an embedded elliptic curve is
concerned, the choice of an ambient space leads to a particular choice
of line bundle. Usually, not all line bundles on the elliptic curve
are restrictions of line bundles on the ambient space; Instead, there
will be some integer $d\in \Z_{>0}$ such that only line bundles
$\Lsheaf$ with $c_1(\Lsheaf) \in d\cdot\Z$ come from the ambient
space. And this integer, called the \emph{degree}, is another measure
for how complicated the ambient space is.  In the remainder of this
section, we will always take $\Lsheaf$ to be a line bundle of minimal
(positive) first Chern class $d$.

The degree is loosely related with how complicated the embedding
is. In the case of a hypersurface in a two-dimensional toric
variety,\footnote{Or: a toric elliptic fibration whose generic ambient
  space fiber is one of the $16$ reflexive polygons.} there are $16$
different ambient spaces corresponding to the $16$ reflexive
polygons. These realize embeddings of degree up to three, the
prototypical examples are~\cite{Braun:2011ux}:
\begin{descriptionlist}
\item[$d=1$:] Long Weierstrass form eq.~\eqref{e:p123} in weighted
  projective space $\CP^2{[1,2,3]}$,
\item[$d=2$:] Hypersurface in $\CP^2{[1,1,2]}$, and
\item[$d=3$:] Cubic in $\CP^2$.
\end{descriptionlist}
If we further consider elliptic curves as complete intersections of
two hypersurface equations in a three-dimensional toric variety, then
there is one additional case:
\begin{descriptionlist}
\item[$d=4$:] Complete intersection of two quadrics in $\CP^3$.
\end{descriptionlist}
The Weierstrass form of the equation (of the Jacobian) can in each
case be derived from the relations between sections of powers of the
minimal line bundle $\Lsheaf$, see~\cite{deligne1975courbes,
  Braun:2014oya}. We have implemented the known
formulas~\cite{MR1858080} in~\cite{Sage}.

However, this does not completely solve the problem of transforming
the toric equation(s) into Weierstrass form. A general formula would
just depend on the coefficients of the defining equations. For the
sake of being explicit, consider a Calabi-Yau hypersurface. Clearly,
we do not need a separate formula for each ambient space: More
constrained hypersurface equations are the result of setting certain
coefficients to zero, corresponding to the embedding of smaller dual
polytopes into larger polytopes. However, already for the case of
hypersurface elliptic curves of degree $d=2$, there are two maximal
dual toric polygons~\cite{Braun:2011ux} (dually, there are two minimal
polygons): $\CP^2[1,1,2]$ and $\CP^1\times \CP^1$. Correspondingly,
there are two different formulas~\cite{pre05771645, Braun:2011ux} for
the Weierstrass form for a toric hypersurface in the degree-$2$ case,
without one being a special case of the other. On the plus side,
though, such an equation can always be derived by looking at a
particular relation between suitable sections of the ``minimal'' line
bundle $\Lsheaf$ and some of its powers, and this is the path we will
take in this paper.

\subsection{Sections of Line Bundles} \label{ss:line_bundles}

Before we derive equations for the relations between line bundles, we
have to discuss how to work with sections in the toric setting. In the
toric hypersurface case, we are familiar with the long exact sequence
of sheaf cohomology when restricting to a divisor (the divisor being
the hypersurface). For a complete intersection $Y\subset X$ of two
equations, that is, sections of $\Osheaf(D_1)$ and $\Osheaf(D_2)$, the
analogous Koszul resolution of the structure sheaf is
\begin{equation}
  \label{eq:Koszul}
  0 \longrightarrow
  \underbrace{\Osheaf_X(-D_1 - D_2)}_{\Rsheaf^{-2}} \longrightarrow
  \underbrace{\Osheaf_X(-D_1) \oplus \Osheaf_X(-D_2)}_{\Rsheaf^{-1}} \longrightarrow
  \underbrace{\Osheaf_X}_{\Rsheaf^0} \longrightarrow 
  \Osheaf_Y \longrightarrow  0.
\end{equation}
A long exact sequence is just a spectral sequence whose $E_1$ tableau
has only two non-zero adjacent columns. Now, we have \emph{three}
columns $q=-2, -1, 0$ in the spectral sequence
\begin{equation}
  \label{eq:SS}
  E_1^{p, q} = 
  H^q(X, \Lsheaf \otimes \Rsheaf^p)
  \quad \Rightarrow \quad
  H^{p+q}(X, \Lsheaf \otimes \Osheaf_Y) = 
  H^{p+q}(Y, \Lsheaf|_Y).
\end{equation}
The first differential $d_1$ is just the induced map of
eq.~\eqref{eq:Koszul} on the sheaf cohomology groups as familiar from
the hypersurface case. However, we now have two new effects to
consider:
\begin{itemize}
\item There are \emph{three} sources for sections of the line bundle
  $\Lsheaf_Y$ restricted to the complete intersection, namely
  \begin{equation}
    \bigoplus_{p} E^{p,-p}_1 =
    H^2(X, \Lsheaf \otimes \Rsheaf^{-2}) \oplus
    H^1(X, \Lsheaf \otimes \Rsheaf^{-1}) \oplus
    H^0(X, \Lsheaf).
  \end{equation}
\item There is a higher differential $d_2:H^1(X,\Lsheaf\otimes
  \Rsheaf^{-2}) \to H^0(X, \Lsheaf)$ that will identify sections of
  $\Lsheaf$ beyond the obvious identifications (coming from $d_1$).
\end{itemize}
The first point is a general problem when studying algebraic varieties
as embedded subvarieties. The sections of a line bundle $\Lsheaf|_Y$
may or may not extend to sections of $\Lsheaf$ over the whole ambient
space $X\supset Y$. If that is not the case, then the choice of
ambient space was an inconvenient one. One should either look for a
different ambient space to embed into, or for a different line bundle
on the ambient space whose sections behave more favorably. As we will
see, in all codimension-two complete intersections there is at least
one favorable line bundle, that is, of low enough degree $\leq 4$ but
with all required sections being induced from the ambient space, such
that we can use it to construct the Weierstrass form.

\subsection{The Second Differential}
\label{sec:d2}

Consider a nef partition $-K = D_1 + D_2$ of the anticanonical divisor
of the three-dimensional ambient toric variety into two numerically
effective divisors $D_1$ and $D_2$. The complete intersection elliptic
curve $Y$ is defined by two polynomials $p_1$, $p_2$ as
\begin{equation}
  Y = V(p_1) \cap V(p_2)
  ,\quad
  p_1 \in H^0(X, D_1)
  ,\quad
  p_2 \in H^0(X, D_2)\,,
\end{equation}
where $V(p)$ denotes the divisor defined by $p=0$.
A section $s$ of a line bundle $\Lsheaf$ always defines a section
$s_Y$ of $\Lsheaf|_Y$ by restriction, but different sections on $X$
might yield the same section on $Y$. Clearly, we can add any section
vanishing on $Y$ to $s$ without changing the restriction. The obvious
candidates of sections of $\Lsheaf$ vanishing on $Y$ are
the image
\begin{equation}
  \label{eq:d1}
  d_1: 
  H^0\big(X, \Lsheaf \otimes \Osheaf(-D_1)\big) + 
  H^0\big(X, \Lsheaf \otimes \Osheaf(-D_2)\big) 
  \xrightarrow{
    \left(\begin{smallmatrix}
      p_1 \\ p_2
    \end{smallmatrix}\right)
  }
  H^0(X, \Lsheaf)   
\end{equation}
Hence, the easy identifications just boil down to working with the
quotient by the image of $d_1$. 

What this section is concerned about is another identification that we
have to perform on the sections on the ambient space, coming from the
$d_2$ differential. To clarify this, we will look at an explicit
example. In fact, the example is very simple. Consider $\CP^1
\times \CP^2$ with the non-product nef partition $D_1 = \Osheaf(1,1)$,
$D_2 = \Osheaf(1,2)$. We let $x_0$, $x_1$ be the two homogeneous
coordinates on $\CP^1$ and $y_0$, $y_1$, $y_2$ be the three
homogeneous coordinates on $\CP^2$. The toric data is also summarized
in \autoref{tab:P1xP2}.
\begin{table}
  \centering
  \begin{tabular}{c|ccccc}
    Homogeneous coordinate & 
    $x_0$ & $x_1$ & 
    $y_0$ & $y_1$ & $y_2$
    \\
    \hline
    Vertex of $\nabla$ &
    \begin{math}\begin{pmatrix}
      1 \\ 0 \\ 0
    \end{pmatrix}\end{math}
    &
    \begin{math}\begin{pmatrix}
      -1 \\ 0 \\0 
    \end{pmatrix}\end{math}
    &
    \begin{math}\begin{pmatrix}
      0 \\ 1 \\ 0
    \end{pmatrix}\end{math}
    &
    \begin{math}\begin{pmatrix}
      0 \\ 0 \\ 1
    \end{pmatrix}\end{math}
    &
    \begin{math}\begin{pmatrix}
      0 \\ -1 \\ -1
    \end{pmatrix}\end{math}
  \end{tabular}
  \caption{The toric variety $\CP^1\times\CP^2$.}
  \label{tab:P1xP2}
\end{table}
A particularly simple choice of equations that nevertheless defines a
smooth complete intersection is
\begin{equation}
  \begin{aligned}
    p_1 =&\; x_0 (y_0+y_1) + x_1 y_2
    &&\in H^0(\CP^1\times\CP^2, D_1)
    \\
    p_2 =&\; x_0 y_2^2 + x_1 y_0 y_1
    &&\in H^0(\CP^1\times\CP^2, D_2).
  \end{aligned}
\end{equation}
We now need to pick a line bundle $\Lsheaf$ on the ambient
$\CP^1\times\CP^2$. The lowest degree choice would be $\Osheaf(1,0)$,
which has degree $2$. However, it has not enough sections on the
ambient space. For example, we would need all four\footnote{A
  degree-$d$ line bundle, $d>0$, on an elliptic curve $Y$ has of
  course $d$ sections.}  sections of
$\Osheaf(1,0)^2|_Y=\Osheaf(2,0)|_Y$ to define the $z$-coordinate in
the Weierstrass model, but $\dim H^0(\CP^1\times\CP^2, \Osheaf(1,0)) =
3$. Hence, we are led to look at the next-smallest degree line bundle
\begin{equation}
  \Lsheaf = \Osheaf(0, 1)
  , \quad
  H^0\big(\CP^1\times\CP^2, \Lsheaf\big)  =
  \Span\{y_0,~ y_1,~ y_2\}
\end{equation}
It is easy to see that the three sections of $\Lsheaf$ restrict to a
basis of three independent sections of $H^0(Y, \Lsheaf|_Y)$ on the
complete intersection. We also remind the reader that the Weierstrass
form in the degree-$3$ case arises as the one relation between the
ten cubic monomials $\Sym^3 H^0(Y, \Lsheaf|_Y)$ inside the
nine-dimensional $H^0(Y, \Lsheaf^3|_Y)$. The first tableau of the spectral
sequence eq.~\eqref{eq:SS} is
\begin{equation}
  E_1^{p, q}(\Lsheaf^3) = 
  H^q(X, \Lsheaf^3 \otimes \Rsheaf^p) =
  \vcenter{ 
    \def\w{16mm} 
    \def\h{8mm} 
    \xymatrix@C=0mm@R=0mm{
      {\scriptstyle q=3} & 
      *=<\w,\h>[F]{ 0 } &
      *=<\w,\h>[F]{ 0 } & 
      *=<\w,\h>[F]{ 0 } 
      \\ {\scriptstyle q=2} & 
      *=<\w,\h>[F]{ 0 } &
      *=<\w,\h>[F]{ 0 } & 
      *=<\w,\h>[F]{ 0 } 
      \\ {\scriptstyle q=1} & 
      *=<\w,\h>[F]{ \C } &
      *=<\w,\h>[F]{ 0 } & 
      *=<\w,\h>[F]{ 0 } 
      \\ {\scriptstyle q=0} & 
      *=<\w,\h>[F]{ 0 } & 
      *=<\w,\h>[F]{ 0 } & 
      *=<\w,\h>[F]{ \C^{10} } 
      \\ & {\scriptstyle p=-2} & {\scriptstyle p=-1} & 
      {\scriptstyle p=0} 
    }} 
  \Rightarrow 
  H^{p+q}(Y, \Lsheaf^3|_Y).
\end{equation}
Clearly, the relation among the ten sections of $H^3(\CP^1\times
\CP^2, \Lsheaf^3)$ is not coming from $d_1$ because the domain
vanishes, see eq.~\eqref{eq:d1}. Instead, we have to quotient by the
image of $d_2$, which is clearly equivalent to knowing the Weierstrass
form of the equation. But we do not know the Weierstrass form yet!
Hence we have to go back to the geometry and use a different approach
to find the relations between the sections.

\subsection{An Algorithm to Compute Relations} \label{ss:algorithm}

Instead, we propose to directly compute the relation between the
sections on the ambient space by restricting to all affine coordinate
patches. Clearly, two sections are equal if they are equal in every
affine patch. In any given patch we can use a local trivialization to
write the sections as polynomials, and polynomials are equal if and
only if their difference is in the ideal generated by the
inhomogenized defining equations. For example, consider the patch $x_1
= y_2 = 1$ in the example of \autoref{sec:d2}. As it turns out, we
only have to consider this single patch in this particular
example. The inhomogenized defining equations define the ideal
\begin{equation}
  I = \langle 
  \hat x_0 (\hat y_0+\hat y_1) + 1,~
  \hat x_0 + \hat y_0 \hat y_1\rangle
  = 
  \langle 
  \hat x_0 \hat y_1^2 - \hat x_0^2 + \hat y_1,~ 
  \hat x_0 \hat y_0 + \hat x_0 \hat y_1 + 1,~ 
  \hat y_0 \hat y_1 + \hat x_0
  \rangle,
\end{equation}
where the second set of generators forms a degrevlex\footnote{That is, a degree reverse lexicographic Gr\"obner basis.}
Gr\"obner basis and we have denoted the inhomogeneous coordinates by hats.
The ten cubics generating $\Sym^3 H^0(Y, \Lsheaf|_Y)$ are, in
inhomogeneous coordinates,
\begin{equation}
  \label{eq:10sections}
  \left\{
    \hat y_0^3,~ \hat y_0^2 \hat y_1,~ \hat y_0 \hat y_1^2,~ \hat y_1^3,~ 
    \hat y_0^2,~ \hat y_0 \hat y_1,~ \hat y_1^2,~ \hat y_0,~ \hat y_1,~ 1
  \right\},
\end{equation}
and their normal form modulo $I$ is 
\begin{equation}
  \left\{
    \hat y_0^3,~ \hat x_0 \hat y_1 + 1,\; -\hat x_0 \hat y_1,~ \hat y_1^3,~ 
    \hat y_0^2,\; -\hat x_0,~ \hat y_1^2,~ \hat y_0,~ \hat y_1,~ 1
  \right\}.
\end{equation}
Hence, the single relation between the ten sections, after
restricting them to the complete intersection and restoring the
homogeneous coordinates, is
\begin{equation}
  y_0^2 y_1 + y_0 y_1^2 - y_2^3 = 0
\end{equation}
This is now the well-known case of a cubic in three homogeneous
variables. Its Weierstrass form is
\begin{equation}
  Y^2 = X^3 + \tfrac{1}{4},
\end{equation}
which has discriminant $\Delta=\tfrac{27}{16}$ and $j$-invariant $0$.

\subsection{Kodaira Map}
\label{sec:kodaira}

We still have considerable freedom in choosing the line bundle
$\Lsheaf$ which realizes the Weierstrass form as the relation between
(powers of) its sections. This is nothing but the Kodaira
map. For example, in the degree-$3$ case the three sections of
$\Lsheaf$ just realize the Kodaira embedding of the elliptic curve $Y$
in $\CP^2$. For the purpose of finding the Weierstrass form, we
want the degree to be as small as possible, and in particular
$\leq 4$. However, as we essentially study the elliptic curve through
its Kodaira map, we can only consider line bundles of positive
degree. Otherwise the Kodaira map would shrink $Y$ to a point, which
obviously would not retain any information. Therefore, a good starting
point for looking for line bundles $\Lsheaf$ on the ambient toric
variety is the cone in $H^2(X,\Z)$ of line bundles with at least one
section. This cone is generated by the first Chern classes of divisors
$V(z_i)$ cut out by a single homogeneous coordinate. The degree on $Y$
is a linear form
\begin{equation}
  \deg( \Lsheaf|_Y) = \int_X D_1 D_2 \; c_1(\Lsheaf),
\end{equation}
so it is just a question of enumerating weighted integer vectors to
list them all up to a certain degree bound.

\subsection{Two Exceptions}
\label{sec:except}

It turns out that there are only two nef partitions (out of $3134$)
for which the above algorithm fails, that is, there is no line bundle
on the ambient toric variety such that
\begin{itemize}
\item The degree $\deg(\Lsheaf|_Y) \leq 4$, and
\item All required\footnote{For degree-$1$, we require the sections of
    $\Lsheaf$, $\Lsheaf^2$, $\Lsheaf^3$, and $\Lsheaf^6$. For
    degree-$2$, we require $\Lsheaf$, $\Lsheaf^2$, and
    $\Lsheaf^4$. For degree-$3$, we require $\Lsheaf$ and
    $\Lsheaf^3$. For degree-$4$, we require $\Lsheaf$ and
    $\Lsheaf^2$.} sections for finding the Weierstrass form are
  restrictions of sections from the ambient space.
\end{itemize}
The two exceptions have the PALP nef ids $(4, 3)$ and $(29, 2)$\footnote{For
an explanation of the notation for the nef ids see
\autoref{ss:nef_partitions}.}. We start with the former,
which is just $\CP^1\times\CP^2$ with the nef
partition $D_1 = \Osheaf(2,1)$ and $D_2 = \Osheaf(0, 2)$. Again using
$[x_0:x_1]\in \CP^1$ and $[y_0:y_1:y_2]\in \CP^2$ as homogeneous
coordinates, the two defining polynomials are
\begin{equation}
  \begin{split}
    p_1 =&\;
    \sum_{i=0}^2
    (a_{00i} x_0^2 + a_{01i} x_0 x_1 + a_{11i} x_1^2) y_i\,,
    \\
    p_2 =&\; 
    \sum_{i,j=0}^2
    b_{ij} y_i y_j = 
    \begin{pmatrix}
      y_0 & y_1 & y_2
    \end{pmatrix}
    \begin{pmatrix}
      b_{00} & b_{10} & b_{20} \\
      b_{01} & b_{11} & b_{21} \\
      b_{02} & b_{12} & b_{22} \\
    \end{pmatrix}
    \begin{pmatrix}
      y_0 \\ y_1 \\ y_2
    \end{pmatrix}\,.
  \end{split}\
\end{equation}
Projection onto the $\CP^1$ factor defines a map $Y=V(\langle p_1, p_2
\rangle) \to \CP^1$. Its pre-image consists of two points: For fixed
$[x_0:x_1] \in \CP^1$, the first equation $p_1$ is a line and the
second equation $p_2$ is a conic in $\CP^2$, which necessarily
intersect in two points. These two points can degenerate to a single
point with multiplicity two, and they must do so at precisely four
pre-images because a torus is the double cover of $\CP^1$ branched at
four branch points. In other words, the discriminant $\delta_{\CP^1}$ of the
double cover $Y\to \CP^1$ is a quartic in the variables $x_0$, $x_1$
with coefficients involving $a$'s and $b$'s but no $y$'s.

The form of the discriminant is constrained by symmetry; $SL(2,\C)
\times SL(3,\C)$ acts naturally on the ambient space. The complete
intersection $Y$ is not invariant under this symmetry, but
its Weierstrass form must be. More formally, we can combine the action
on the homogeneous coordinates with an action on the coefficients such
that the combined action does not change the equations $p_1$,
$p_2$. For example, the $M_3\in SL(3,\C)$-part of the action is
\begin{equation}
    \begin{pmatrix}
      y_0 \\ y_1 \\ y_2
    \end{pmatrix}
    \mapsto
    M_3 
    \begin{pmatrix}
      y_0 \\ y_1 \\ y_2
    \end{pmatrix}
    ,\quad
    \begin{pmatrix}
      a_{ij0} \\  a_{ij1} \\  a_{ij2} \\ 
    \end{pmatrix}
    \mapsto
    M_3^{-1} 
    \begin{pmatrix}
      a_{ij0} \\  a_{ij1} \\  a_{ij2} \\ 
    \end{pmatrix}
    ,\quad
    (b_{ij}) \mapsto (M_3^{-1})^T (b_{ij}) M_3^{-1}.
\end{equation}
A \emph{covariant} is a polynomial that does not transform under the
combined group action, obvious examples are $p_1$ and $p_2$. An
\emph{invariant} is a covariant that, furthermore, does not depend on
the homogeneous coordinates, for example $\det(b_{ij})$. The
discriminant $\delta_1$ that we are looking for must be a covariant of
bi-degree $(4,0)$ in $[x_0:x_1]$ and $[y_0:y_1:y_2]$.

The tersest way to characterize $\delta_1$ completely is as the
$\Theta'$-invariant~\cite{SalmonConics, trac17305} of the system of
two conics $(p_1^2, p_2)$. That is, ignore the action on the $\CP^1$
factor for the moment and consider $p_1^2$ and $p_2$ as two quadratics
in $[y_0:y_1:y_2]$. The determinant $\Delta$ of the coefficient matrix
of a quadratic is clearly an invariant of the action on $\CP^2$, hence
so is every $\epsilon$-coefficient in the formal
expansion\footnote{The invariants $\Delta(p_1^2)$ and $\Theta(p_1^2,
  p_2)$ vanish because $p_1^2$ is a degenerate conic.}
\begin{equation}
  \Delta(p_1^2 + \epsilon p_2) = 
  \Delta(p_1^2) + 
  \epsilon \Theta(p_1^2, p_2) +
  \epsilon^2 \Theta'(p_1^2, p_2) +
  \epsilon^3 \Delta(p_2)
\end{equation}
We note that $\delta_1(x_0,x_1) = \Theta'(p_1^2, p_2)$ is quartic in
$x_0$ and $x_1$, quadratic in the coefficients $a_{ijk}$ and quadratic
in the coefficients $b_{ij}$. Finally, the equation of a double cover
branched at the zeroes of $\delta_1$ is
\begin{equation}
  Y^2 = \delta_1(x_0, x_1),
\end{equation}
for which we already know how to write the Weierstrass
form~\cite{MR1858080, Sage}.

It remains to consider the second exceptional case, that is, the one
with PALP nef id $(29, 2)$. Geometrically, it is the product
$\CP^1\times dP_1$, that is, a simple blowup\footnote{We use the
  notation where $\CP^2=dP_0$.} of the first case along a curve
$\CP^1\times \ptset$. Moreover, the two divisors defining the nef
partition are just the pull-backs of the two divisors of the first
case. In terms of toric geometry, this means that the dual polytope
$\nabla$ contains the dual polytope of $\CP^1\times\CP^2$. Dually, the
polytope $\Delta$ is contained in the polytope of
$\CP^1\times\CP^2$. Hence the formula for bringing the complete
intersection into Weierstrass form is simply a specialization of the
formula from the first case where some coefficients are set to zero.


 \section{Classifying Toric Mordell-Weil Groups} 
\label{s:classification}

We begin this section by reviewing how to construct Calabi-Yau manifolds as complete
intersections in toric varieties. Having laid the general groundwork, we then calculate
all nef partitions of three-dimensional reflexive polytopes and give a short summary
of our results. Next, we recall the concept of \emph{toric} Mordell-Weil groups
as introduced in \cite{Braun:2013nqa} and explain how to compute them for a given
ambient fiber space.
Finally, we determine the toric Mordell-Weil group for every elliptic fiber
embedded in a three-dimensional toric variety corresponding to a reflexive polytope
and comment on our results.

\subsection{Complete Intersections in Toric Varieties}

As discovered by Batyrev \cite{Batyrev:1994hm,Batyrev:1994pg}, toric geometry provides a convenient way of constructing
Calabi-Yau manifolds embedded in ambient toric varieties either as hypersurfaces
or as complete intersections. Conveniently, Batyrev's construction is combinatorial:
Given a lattice polytope $\Delta$ in a lattice $N \simeq \mathbb{Z}^{n+1}$, its dual (or polar)
polytope is given by
\begin{align}
 \Delta^\circ := \{ y \in M | \langle x, y \rangle  \geq -1 \ \forall x \in \Delta \}\,.
\end{align}
Here $M$ is the dual lattice of $N$.
If $\Delta^\circ$ is again a lattice polytope, then $\Delta$ is called \emph{reflexive}. Furthermore,
since $(\Delta^\circ)^\circ = \Delta$, $\Delta^\circ$ is reflexive if and only if $\Delta$ is reflexive.
Next, we take all lattice points of $\Delta^\circ$ that are not interior points of a facet\footnote{That is,
a face of codimension one.} to construct a fan from a fine star triangulation of these points with respect
to the origin and call the corresponding toric variety $X_{n+1}$. Denote the homogeneous coordinates of
$X_{n+1}$ by $z_i$ and the respective points of $\Delta^\circ$ by $x_i$.
Consider then the hypersurface $Y_n$ inside $X_{n+1}$ given by the equation
\begin{align} \label{e:hypersurface}
 p = \sum_{y_j \in \Delta} a_j \prod_{i} z_i^{\langle y_j, x_i \rangle + 1}\,.
\end{align}
It defines a Calabi-Yau $n$-fold inside $X_{n+1}$ and there exist simple combinatorial formulas
in terms of the data of $\Delta$ and $\Delta^\circ$ to compute its cohomology dimensions.
Furthermore, it is worth to note that by exchanging $\Delta$ and $\Delta^\circ$ one 
obtains the mirror manifold of $Y_n$.

To generalize this approach to complete intersections, one must specify additional information.
In the hypersurface case, the homology class of the divisor defined by the vanishing of 
\eqref{e:hypersurface} must be Poincar\'e-dual to the cohomology class of the first Chern class of
the ambient space in order for the hypersurface to be Calabi-Yau. If instead the Calabi-Yau manifold
is to be the intersection of several divisors, then their sum must still be dual to the first Chern
class of the ambient space. However, the classes of the individual divisors are not fixed anymore.

One such way of additionally specifying the classes of the divisors defining the complete
intersection proceeds by giving a \emph{nef partition} of the reflexive polytope $\Delta^\circ$. A nef partition
of $\Delta^\circ$ into $r$ parts is a set of lattices polytopes $\Delta_i$ and $\nabla_i$ with $i=1,\dots,r$
satisfying
\begin{align}
 \Delta &= \Delta_1 + \dots + \Delta_r & \Delta^\circ &= \langle \nabla_1, \dots, \nabla_r \rangle_{\textrm{conv}} \nonumber \\
 \nabla^\circ &= \langle \Delta_1, \dots, \Delta_r \rangle_{\textrm{conv}} & \nabla &= \nabla_1 + \dots + \nabla_r
\end{align}
with $\langle \cdot, \dots, \cdot \rangle_{\textrm{conv}}$ the convex hull, $+$ Minkowski addition, and
\begin{align}
  \left(\nabla_n, \Delta_m \right) &\geq - \delta_{nm}\,,
\end{align}
where here we mean this to hold for every pair of points from $\nabla_n$ and $\Delta_m$. Effectively,
we have split the vertices of $\Delta^\circ$ into $r$ disjoint subsets spanning the polytopes
$\nabla_i$ and made sure that they fulfill certain additional constraints.
Given such
a nef partition, we again define $X_{n+r}$ to be the ambient variety obtained from $\Delta^\circ$ as
above. Furthermore, the nef partition specifies the following $r$ equations defining the Calabi-Yau manifold
$Y_n$:
\begin{align}
 p_m &= \sum_{y_j \in \Delta_m} a_{m, j} \prod_{n=1}^r \prod_{x_i \in \nabla_n} z_i^{\langle y_j, x_i \rangle +\delta_{nm}}\,,
 \qquad m=1,\dots,r\,.
\end{align}
Note that one can also interpret a nef partition of $\Delta^\circ$ as a nef partition of $\nabla^\circ$. In doing
so, one exchanges $Y_n$ by its mirror. Let us point out that the ambient space of a mirror manifold can differ
for different nef partitions of the same polytope.

Finally, we remark that there are two special cases of nef partitions. The simplest one corresponds to
\emph{direct products}. Given nef partitions of two reflexive polytopes ${\Delta^{(1)}}^\circ$
and ${\Delta^{(2)}}^\circ$, these define a nef partition of the polytope $\Delta^{(1)} \times \Delta^{(2)}$.
The corresponding complete intersection manifold is then a direct product of complete intersections inside
the direct product of the varieties corresponding to ${\Delta^{(1)}}^\circ$ and ${\Delta^{(2)}}^\circ$.
The other special case corresponds to \emph{projections}. If a nef partition has one component $\nabla_i$ that
is spanned only by a single vertex $v$, then the complete intersection can be reduced to a complete intersection
in a toric variety of one dimension less whose reflexive polytope is obtained by projecting $\Delta^\circ$
along $v$.

\subsection{Nef Partitions of 3d Lattice Polytopes} 
\label{ss:nef_partitions}

As a test sample for applying our Weierstrass algorithm we use elliptic curves that are embedded in three-dimensional
toric varieties and we therefore spend a moment to construct the corresponding nef partitions.
It is well-known that the number of reflexive polytopes of a given dimension is finite,
but increases very quickly with the dimension: In two dimensions, there are precisely $16$ reflexive polygons,
in three dimensions there exist $4319$ reflexive polytopes \cite{Kreuzer:1998vb}, and the 
$473,800,776$ reflexive polytopes in four dimensions were determined in \cite{Kreuzer:2000xy}.
The exact number in five dimensions is unknown, but expected to be large enough to currently make
its computation unfeasible.
\begin{figure}[h]
 \includegraphics[width=0.95\textwidth]{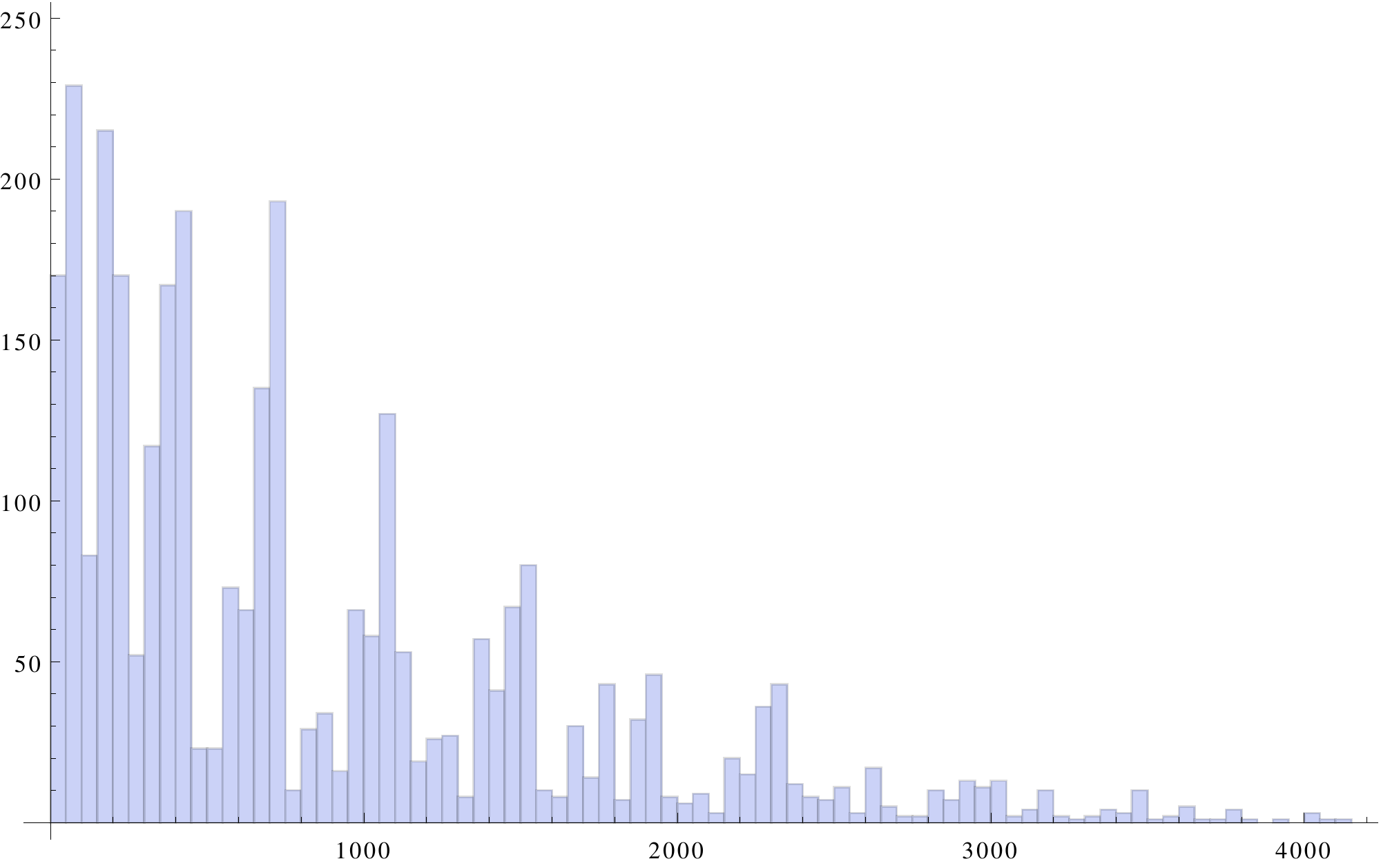}
 \centering
 \label{f:nef_distribution}
 \caption{Histogram of the number of nef partitions of the $4319$ reflexive polytopes in three dimensions.}
\end{figure}
\begin{figure}[h]
\centering
 \includegraphics[width=0.95\textwidth]{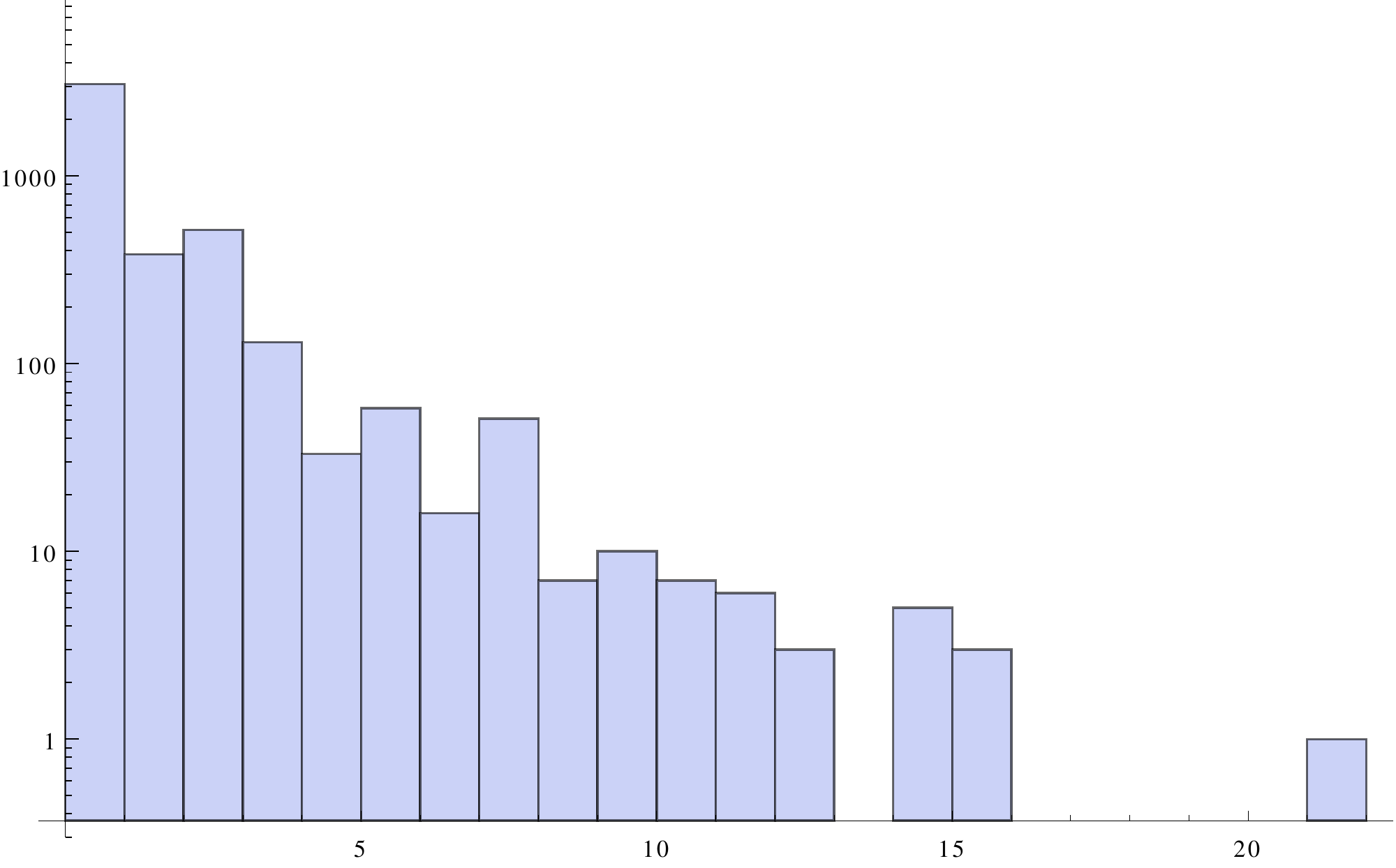}
 \label{f:number_nef_distribution}
 \caption[Histogram of the number of polytopes that have a given number of nef partitions.]{Histogram
 of the number of polytopes that have a given number of nef partitions. There are $3090$
 reflexive three-dimensional polytopes that do not admit a nef partition. The reflexive polytope with PALP id
 $214$ has the most nef partitions, namely $21$.}
\end{figure}
In the case of the $4319$ three-dimensional polytopes, the nef partitions can be computed using PALP \cite{Kreuzer:2002uu} 
via Sage \cite{Sage} within a matter of minutes. One finds that there exist $3134$ nef partitions. $16$ of these
correspond to direct products embedded in $F_i \times \mathbb{P}^1$ for the $16$ two-dimensional varieties,
and $807$ correspond to projections.

Last but not least, let us introduce a nomenclature for denoting the nef partitions dealt with in the following
subsections. Three-dimensional reflexive polytopes already have a unique id as assigned by the PALP database.
This id obeys
\begin{align}
  \#_{\textrm{points}}(P) < \#_\textrm{points}(P') \quad \Rightarrow \quad id(P) < id(P')
\end{align}
and
\begin{align}
 \#_{\textrm{points}}(P) = \#_\textrm{points}(P')\ \wedge \  \#_{\textrm{vertices}}(P) < \#_\textrm{vertices}(P') \quad \Rightarrow \quad id(P) < id(P')\,,
\end{align}
that is, the polytopes are ordered by the number of integral points and the number of vertices. Sage can be used to compute the PALP index
of a given reflexive polytope. To furthermore identify the nef partitions uniquely,
we run $\texttt{nef.x}$  via the
\begin{equation}
 \texttt{ReflexivePolytope.nef\_partitions()}
\end{equation}
method of Sage on a given reflexive
polytope in PALP normal form. This output is uniquely ordered and allows us to assign ids to the different nef partitions.
By a nef partition with id $(i, j)$ we therefore mean the $(j + 1)$th nef partition of the three-dimensional reflexive polytope with
PALP id $i$ as determined by the \texttt{nef\_partitions()} method of Sage.

\subsection{Toric Mordell-Weil Groups}

Next, we introduce the concept of \emph{toric} Mordell-Weil groups of an elliptic fiber. First however,
let us quickly recall a few facts about elliptic curves. An elliptic curve is a genus-one curve, i.e. a $T^2$,
together with one special marked point that defines the zero point of the curve.
Given such an elliptic curve $E(K)$ over some field $K$, it is well-known that the set of
points on this elliptic curve with coefficients in $K$ forms a group, called the Mordell-Weil group $\MW(E)$ of the curve.
The group action can easily be understood visually: In order to add two points $P$ and $Q$, intersect the elliptic curve $E$
with the line passing through both $P$ and $Q$. It is guaranteed to have a third intersection with $E$, which we denote by
$R$. Construct another line passing through $R$ and the zero point of the elliptic curve. The third intersection point
of this line will be $P+Q$. 
While it is straightforward to show that this does indeed define a valid Abelian group action\footnote{For special cases, a proof
and expressions in coordinate form see for example \cite{deligne1975courbes,silverman2009arithmetic}.}, it is a highly non-trivial fact that the Mordell-Weil group
is finitely generated.

Now we would like to consider fibrations $Y_n$ of elliptic curves over base manifolds $B_{n-1}$. Non-trivial fibrations of this kind
imply that the complex structure of the elliptic curve varies from point to point in the base and, equivalently, one
can view such a fibration as an elliptic curve over the field of rational functions on the base manifold.
With respect to this function field  the rational points of the elliptic curve correspond to the global sections
\begin{align}
 f_i: B_{n-1} \rightarrow Y_n
\end{align}
of the fibration. In particular, for a non-singular elliptic fibration
one has the relation
\begin{align} \label{eq:h11_mwrank}
 h^{1,1}(Y_n) = h^{1,1}(B_{n-1}) + \rk \MW(Y_n) + 1\,.
\end{align}
Here the $+1$ is owed to the fact that it takes $n+1$ independent
global sections $f_1,\dots,f_{n+1}$ in order
to generate a Mordell-Weil group of rank $n$, 
since one section must serve as the zero point, or neutral element, of the elliptic fiber. If one takes
$f_0$ as zero section, then
\begin{align}
 \sigma_i := f_i - f_0
\end{align}
can be used as generators of the Mordell-Weil group.

Given a general elliptic fibration, it is a difficult problem to determine all global sections, even though their
total number can be computed using \eqref{eq:h11_mwrank} and generalizations thereof. In particular, there
exist examples for which the homology classes of the sections can be determined, but their precise coordinate expressions
cannot \cite{Braun:2013yti}. More importantly, the total Mordell-Weil group generally depends on the entire fibration
and can therefore not be computed independently of the base.
Nevertheless, there exists a subgroup of the Mordell-Weil group, the \emph{toric} Mordell-Weil group, that indeed depends
only on the toric variety the elliptic fiber is embedded in and can therefore be computed without reference to
a specific base manifold or fibration. Let us therefore explain how the toric Mordell-Weil group is defined by reviewing
the material of \cite{Braun:2013nqa}.

Denote the toric ambient fiber space by $W_{1+c}$, where $c$ is the codimension of the elliptic fiber $E$. 
Then the homogeneous coordinates $z_i$ of $W_{1+c}$ define toric divisors $V(z_i)$ given by the vanishing of a single
homogeneous coordinate. If such a divisor intersects the elliptic curve once, i.e. is satisfies
\begin{align}
 \int_{E} V(z_i) = 1\,,
\end{align}
then this divisor will become a global section of the fibration after fibering $W_{1+c}$ over the base manifold.
We call these divisors the \emph{toric} global sections and call the subgroup
\begin{align}
 \MW_T(E) \subseteq \MW(E)
\end{align}
the toric Mordell-Weil group. In \cite{Braun:2013nqa} the toric Mordell-Weil groups of elliptic curves
embedded as hypersurfaces inside two-dimensional toric varieties were analyzed. In the next subsection,
we will apply the same analysis using the new algorithm for Weierstrass forms developed in \autoref{s:algorithm}.

\subsection{Results for Elliptic Curves of Codimension Two}

In the final subsection of this chapter, we present the main results of our computations.
Before proceeding to the results, let us remark on how to compute the Mordell-Weil group laws for
a given fibration in practice. While we computed the Weierstrass forms of the elliptic curves by keeping the coefficients
in the complete intersection equations general, this approach makes little sense for determining the Mordell-Weil group laws.
Instead, we generated a considerable number\footnote{By considerable, we mean $\mathcal{O}(100)$ in order to make
sure that we indeed obtain a generic example.} of curves with random complex structure coefficients in $\mathbb{Z}$.
We then computed the explicit coefficients of the points cut out by toric sections, mapped these
to the elliptic curve in Weierstrass form and determined the relations between them.
Special care has to be taken when mapping the points from the original elliptic curve to the curve in Weierstrass form.
As discussed in \autoref{s:algorithm} our map works through an intermediate embedding inside $\mathbb{P}_{231}$,
$\mathbb{P}_{112}$, $\mathbb{P}^2$, or $\mathbb{P}^3$. However, the maps from the last three spaces to Weierstrass form
are not injective: They in fact map the elliptic curves $4:1$, $9:1$ and $16:1$, respectively. As a consequence, distinct
points on the original curve may be mapped to the same point of the curve in Weierstrass form and therefore
torsion factors of the Mordell-Weil group may get lost. To make sure that we find the correct
torsion groups, it is therefore crucial to use different embeddings of the same curve in case that the points
on the curve in Weierstrass satisfy non-trivial relations with respect to the Mordell-Weil group law.
While the map from $\mathbb{P}^2$ to Weierstrass
form may eliminate a $\mathbb{Z}_3$ torsion factor, the map from $\mathbb{P}_{112}$ will not,
and one can therefore determine the correct toric Mordell-Weil groups even in the presence of torsion.

The computations were performed using PALP \cite{Kreuzer:2002uu}, Sage \cite{Sage}
and in particular the Sage modules for polytopes \cite{sage_polytope} and toric geometry \cite{sage_toric}. Furthermore,
we made heavy use of the Sage interface to Singular \cite{DGPS}.
For every nef partition of a reflexive three-dimensional polytope $\Delta^\circ$, we computed the following data:
\begin{itemize}
 \item The two defining equations of the complete intersection with general coefficients $a_i$.
 \item The Weierstrass coefficients $f$ and $g$ of equation \eqref{e:wf} in terms of $a_i$.
 \item The integral points $v_i$ of $\Delta^\circ$ that are promoted to toric sections $V(z_i)$
       after fibering the elliptic curve over a base manifold.
 \item The relations between the Mordell-Weil generators $\sigma_i$ after choosing a zero point on the elliptic curve.
 \item The resulting toric Mordell-Weil group, including its torsion part.
 \item The Kodaira types of the non-toric singularities that occur if all $a_i$ are generic.
\end{itemize}

Since the full list of results is too long to be included in the text of this paper, we have created a website at
\begin{align}
  \textrm{\url{http://wwwth.mpp.mpg.de/members/jkeitel/Weierstrass/}}
\end{align}
with a database of the $3134$ nef partitions of three-dimensional reflexive polyhedra. For each such nef partition,
there exists a file of the form \texttt{RP\_NEF.txt}. Let us illustrate the file format using the nef
partition $(2355, 0)$:
\begin{verbatim}
Summary for nef partition with id (2355, 0).

Defining data of the nef partition:
rays = [z0: (1, 0, 0), z1: (0, 1, 0), z2: (0, 0, 1), z3: (-1, 1, 1),
z4: (2, -1, -1), z5: (1, 0, -1), z6: (1, -1, 0), z7: (-1, 1, 0),
z8: (-1, 0, 1), z9: (-2, 1, 1), z10: (1, -1, -1), z11: (0, 0, -1),
z12: (0, -1, 0), z13: (-1, 0, 0)]
nabla_1 = (0, 1, 2, 3, 4, 5, 6)
nabla_2 = (7, 8, 9, 10, 11, 12, 13)

Toric Mordell-Weil group:
zero = (0, 1, 0)
generators = [s0: (0, 0, 1), s1: (2, -1, -1), s2: (-2, 1, 1),
s3: (0, 0, -1), s4: (0, -1, 0)]
relations = [s0-s3 = (1), s1-s2 = (1), s4 = (1)]
group = Z^2 x Z_2

Complete intersection equations:
p1 = a3*z0*z1*z2*z3*z4*z5*z6 + a2*z1*z3*z5*z7*z9*z11*z13
+ a1*z2*z3*z6*z8*z9*z12*z13 + a0*z4*z5*z6*z10*z11*z12*z13
p2 = a7*z0*z1*z2*z3*z7*z8*z9 + a6*z0*z1*z4*z5*z7*z10*z11
+ a5*z0*z2*z4*z6*z8*z10*z12 + a4*z7*z8*z9*z10*z11*z12*z13

Weierstrass coefficients:
f = [...]
g = [...]

Generic non-Abelian singularities:
a7: (0, 0, 2), I_2
a6: (0, 0, 2), I_2
a5: (0, 0, 2), I_2
a4: (0, 0, 2), I_2
a3: (0, 0, 2), I_2
a2: (0, 0, 2), I_2
a1: (0, 0, 2), I_2
a0: (0, 0, 2), I_2
\end{verbatim}

The first block summarizes the toric data defining the nef partition. The first line defines the variable names
$z_i$ assigned to the homogeneous variables associated with each ray of the ambient fan and the second line
specifies the nef partition by listing the indices of the rays spanning $\nabla_1$ and $\nabla_2$. In this
example
\begin{align}
  \nabla_1 = \langle v_0 v_1 v_2 v_3 v_4 v_5 v_6 \rangle_{\textrm{conv}}\,, \qquad
  \nabla_2 = \langle v_7 v_8 v_9 v_{10} v_{11} v_{12} v_{13} \rangle_{\textrm{conv}}\,.
\end{align}
The second paragraph contains information about the toric Mordell-Weil group. This particular example has six divisors
that become (not necessarily independent) sections after fibering the elliptic curve over a base manifold and the toric
Mordell-Weil group generated by these divisors is $\mathbb{Z}^2 \oplus \mathbb{Z}_2$.
Choosing the divisor corresponding to the ray $\begin{pmatrix}0&1&0\end{pmatrix}^T$ as the divisor that cuts out the neutral
element on the curve, the remaining five divisors $\sigma_i$, $i=0,\ldots,4$ satisfy three relations. To specify these relations
we denote by $(i)$ the generator of the torsion part times $i$. Here, this means that the section $\sigma_4$ generates the
$\mathbb{Z}_2$ factor and, up to this torsion part, the pairs of sections $\sigma_0$ and $\sigma_3$, and $\sigma_1$ and $\sigma_2$, 
are identified under the Mordell-Weil group law.
Next, the record contains the two complete intersection equations in order to define the coefficients $a_i$ determining
the complex structure of the elliptic curve. The Weierstrass coefficients (omitted here due to their length) are then
given in terms of the $a_i$.
Finally, we list the non-Abelian singularities that a such an elliptic curve with generically chosen $a_i$ will have. In this case,
there is an additional $SU(2)^8$ gauge group with branes located along the eight base loci $a_i=0$ for $i=0, \ldots 7$.

\subsubsection*{Statistics of the $3134$ elliptic curves of codimension two}
Let us give a quick summary of the results we found. We begin by noting that $16$ of the $3134$
nef partitions are direct products. Up to lattice isomorphisms, they are obtained as
\begin{align}
 \nabla_1 = \langle \begin{pmatrix} 1\\0\\0\end{pmatrix}, \begin{pmatrix} -1\\0\\0\end{pmatrix} \rangle_{\textrm{conv}}\,, \qquad
 \nabla_2 = \langle \begin{pmatrix} 0\\ v_i\end{pmatrix} \textrm{ where } v_i \in F_j \rangle_{\textrm{conv}} \,,
\end{align}
where $F_j$ is one of the $16$ reflexive polygons. Their PALP ids are contained in \autoref{t:products}.
The total ambient space corresponding to the face fan of $\Delta^\circ$ is
$\mathbb{P}^1 \times F_j$ and the complete intersection factors into a quadratic equation inside $\mathbb{P}^1$ and the anticanonical
hypersurface in $F_j$. Therefore these nef partitions consist of two disjoint elliptic curves, each of which is described by
a hypersurface inside a two-dimensional toric variety. Both of them have the same complex structure.
\begin{table}[h]
 \centering
\begin{tabular}{c|cccccccc}
 $\mathbb{P}^1 \times$ & $F_1$ & $F_2$ & $F_3$ & $F_4$ & $F_5$ & $F_6$ & $F_7$ & $F_8$\\
 \hline
 PALP id & $(4, 2)$ & $(30,1)$ & $(29,3)$ & $(17, 1)$ & $(84,8)$ & $(61,2)$ & $(218,0)$ & $(149,3)$
\end{tabular}
\\
\vspace{0.5cm}

\begin{tabular}{c|cccccccc}
 $\mathbb{P}^1 \times$ & $F_9$ & $F_{10}$ & $F_{11}$ & $F_{12}$ & $F_{13}$ & $F_{14}$ & $F_{15}$ & $F_{16}$\\
 \hline
 PALP id & $(194, 5)$ & $(113, 0)$ & $(283, 0)$ & $(356, 3)$ & $(453, 0)$ & $(505, 0)$ & $(509, 0)$ & $(768, 1)$
\end{tabular}
  \caption{The PALP ids for the $16$ nef partitions that are direct products inside the spaces $\mathbb{P}^1 \times F_i$, where $F_i$
  is a reflexive polygon.}
  \label{t:products}
\end{table}
Clearly, set-ups of this kind do not occur in F-theory compactifications with fibers defined as
hypersurfaces. It would be interesting to study the resulting low-energy effective theories of such compactifications further,
but we reserve this for future work. As these spaces appear to make up a class of their own, we will not include them
in our analyses below and instead restrict to the remaining $3118$ nef partitions.

\begin{figure}[h]
\centering
\begin{minipage}{0.48\textwidth} \centering
 \includegraphics[width=0.9\textwidth]{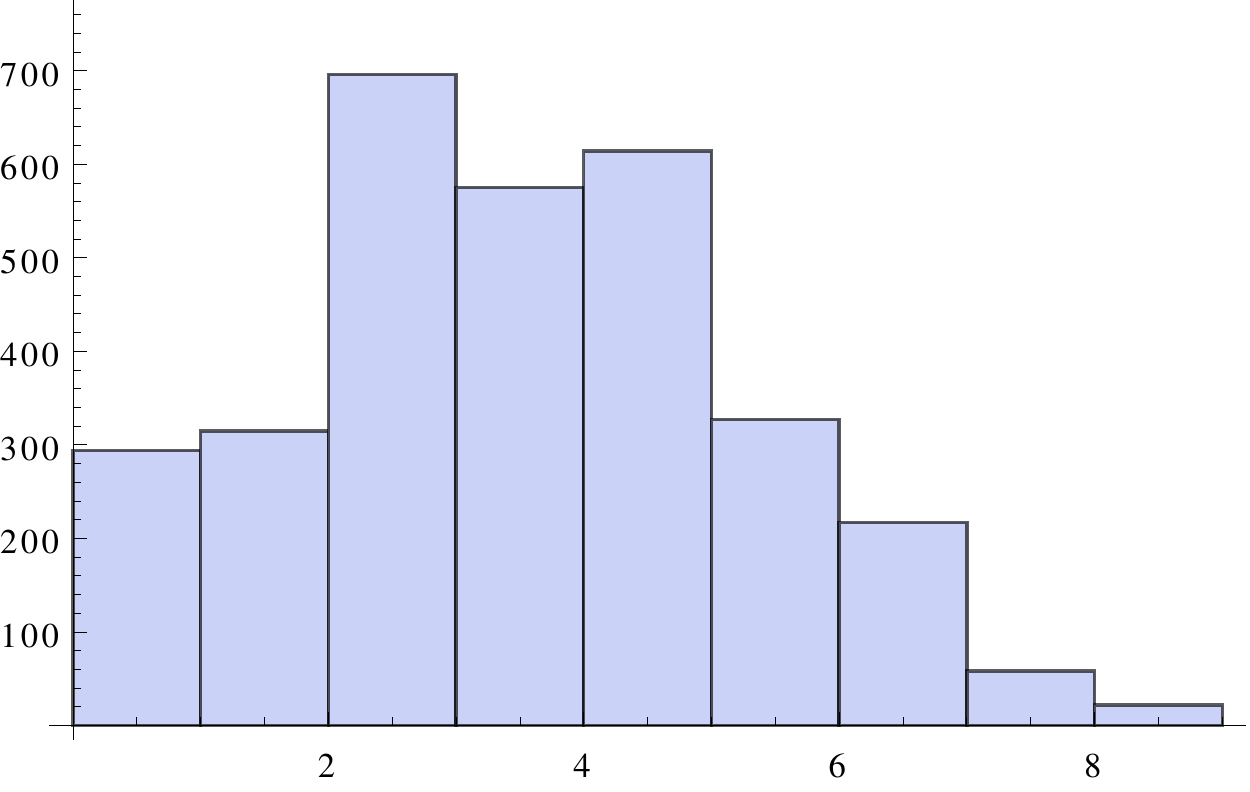}
\end{minipage}%
\begin{minipage}{0.48\textwidth} \centering
 \begin{tabular}{c|c}
   Manifolds & Toric sections\\
   \hline
   294 & 0 \\
   315 & 1 \\
   696 & 2 \\
   575 & 3 \\
   614 & 4 \\
   327 & 5 \\
   217 & 6 \\
   58 & 7 \\
   22 & 8
  \end{tabular}
\end{minipage}
 \caption{Histogram of the number of toric sections for the $3118$ nef partitions of three-dimensional reflexive polytopes that are not direct products.}
  \label{f:sections}
\end{figure}
\begin{figure}[h]
\centering
\begin{minipage}{0.48\textwidth} \centering
 \includegraphics[width=0.9\textwidth]{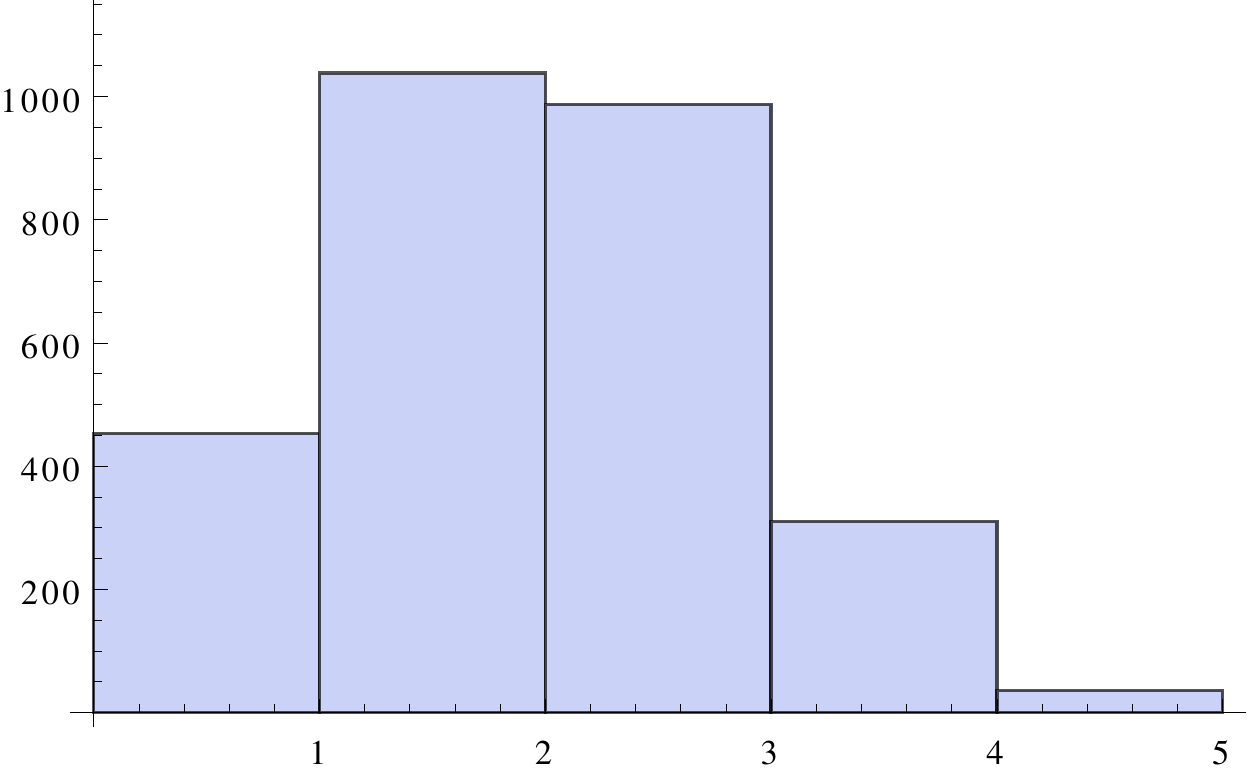}
 \end{minipage}%
\begin{minipage}{0.48\textwidth} \centering
 \begin{tabular}{c|c}
   Manifolds & Mordell-Weil rank\\
   \hline
   453 & 0 \\
   1038 & 1 \\
   987 & 2 \\
   310 & 3 \\
   36 & 4 \\
  \end{tabular}
\end{minipage} 
 \caption[Histogram of the toric Mordell-Weil rank for the nef partitions of three-dimensional reflexive polytopes.]{Histogram
 of the toric Mordell-Weil rank for the nef partitions of three-dimensional reflexive polytopes. The $326$
 complete intersections that are either a direct product or do not have a toric section are excluded.}
  \label{f:ranks}
\end{figure}

We list in \autoref{f:sections} the distribution
of the number of toric divisors corresponding to sections among the complete intersection curves. Note that not all
of these divisors will be independent in homology. In \autoref{f:ranks} we give the distribution of the toric Mordell-Weil
ranks. The highest \emph{toric} rank that we find is four. 
\begin{table}[h]
 \begin{tabular}{ccccccccccc}
   Trivial group & $\mathbb{Z}_2$ & $\mathbb{Z}_3$ & $\mathbb{Z}_4$ & $\mathbb{Z}$ & $\mathbb{Z} \oplus \mathbb{Z}_2$ &
  $\mathbb{Z}^{\oplus 2}$ & $\mathbb{Z}^{\oplus 2} \oplus \mathbb{Z}_2$ & $\mathbb{Z}^{\oplus 3}$ &
  $\mathbb{Z}^{\oplus 3} \oplus \mathbb{Z}_2$ & $\mathbb{Z}^{\oplus 4}$ \\
  \hline
  $315$ & $113$ & $24$ & $1$ & $931$ & $107$ & $985$ & $2$ & $309$ & $1$ & $36$
 \end{tabular}
 \caption[Toric Mordell-Weil groups for the elliptic fibers of codimension two.]{The
 full toric Mordell-Weil groups for the elliptic fibers of codimension two. Note that we have omitted direct products and the genus-one
 curves that do not have a single toric point.}
 \label{t:mw_groups}
\end{table}
Naturally, not all groups of the same rank are equal, as some
have additional torsion factors. In \autoref{t:mw_groups} we give a complete survey of the toric Mordell-Weil groups for the models
that possess at least one toric section.
As one might expect, there are additional toric Mordell-Weil groups when compared with the elliptic curves that are embedded
in toric surfaces. The groups that do not occur for elliptic curves that are hypersurfaces are
\begin{align}
 \mathbb{Z}_4,\quad \mathbb{Z}^{\oplus 2} \oplus \mathbb{Z}_2,\quad \mathbb{Z}^{\oplus 3} \oplus \mathbb{Z}_2,\quad \mathbb{Z}^{\oplus 4}\,.
\end{align}

Last, but not least, let us comment on the appearance of generic non-Abelian gauge groups. As noted in \cite{Braun:2013nqa} and
recently examined in detail in \cite{Klevers:2014bqa}, certain fibers can generically induce non-Abelian singularities.
These generic non-Abelian singularities differ from the ones induced by tops \cite{Candelas:1996su,Candelas:2012uu}.
When a non-Abelian singularity is enforced by a top, the ambient space of the elliptic fiber becomes
reducible over a divisor in the base and as a consequence, the elliptic fiber does, too. In the
case of these generic non-Abelian singularities the ambient space remains irreducible, but the fiber splits into various irreducible pieces.
Such non-Abelian singularities cannot be read off directly from the toric data of the ambient space and therefore we called
them \emph{non-toric} non-Abelian singularities in \cite{Braun:2013nqa}. Note also that the base locus over which such singularities occur
is \emph{not} defined by the vanishing of a single homogeneous coordinate, but rather a polynomial in the base coordinates.

Since these non-toric singularities are not directly visible in the defining data of the ambient space,
the exceptional divisors do not belong to rays of a top, but instead to rays that are part of the fan defining the ambient space of the generic
fiber. Since the maximum number of integral points of a reflexive polytope of given dimension is bounded from above, the maximum
number of non-toric exceptional divisors and therefore the total rank of the non-toric gauge group is, too. To illustrate this,
consider the $16$ reflexive polygons. $F_{16}$\footnote{Here we are using the notation of \cite{Braun:2013nqa}, in which
$F_{16} \cong \mathbb{P}^2/\mathbb{Z}_3$} is the one with most integral
points, namely ten. The nine non-zero points give rise to seven independent homology classes. One of them corresponds to the neutral
element of the elliptic curve, so the maximum allowed gauge rank is six. In fact, one can show that
the maximal non-toric gauge group is $SU(3)^3 / \mathbb{Z}_3$ \cite{Klevers:2014bqa}.

Since three-dimensional reflexive polytopes can contain more integral points than their two-dimensional analogues (the largest one
has $39$ integral points), the non-toric gauge group content is considerably more diverse. Not only can one find non-toric GUT candidates,
but there are also fibers that generically exhibit $E_6$, $E_7$, and $E_8$ singularities. In \autoref{a:groups} we list
the non-toric singularities for the $3118$ non-product nef partitions.


 \section{Examples} \label{s:examples}
Having studied the toric Mordell-Weil groups of the elliptic curves of codimension two,
the next natural step would be to classify their \emph{tops}, i.e. all ways of
generating non-Abelian singularities torically. While the classification of two-dimensional
tops was achieved in \cite{Bouchard:2003bu},
three-dimensional tops have so far not been studied. 
However, as these tops appear to have a fairly involved structure, we reserve this
task for future work. Instead, we present several interesting examples
illustrating features that do not occur for fibers in toric surfaces.

\subsection
[\texorpdfstring{$SU(5) \times U(1)^2$}{SU(5)xU(1)xU(1)} with Different Antisymmetric Representations]
{\boldmath $SU(5) \times U(1)^2$ with Different Antisymmetric Representations}

Let us begin with the example that motivated this work in the first place: An $SU(5)$ GUT model with $U(1)$ factors.
As mentioned in the introduction, \emph{fully resolved} $SU(5)$ F-theory models with fibers embedded as hypersurfaces
suffer from the constraint that their antisymmetric representations always have the same
charge under additional $U(1)$ gauge factors. For complete intersection fibers,
we do not expect this to happen anymore.

\begin{table}[h]
 \centering
 \begin{tabular}{cccccc}
  $v_0$ & $v_1$ & $v_2$ & $v_3$ & $v_4$ & $v_5$\\
  \hline
  $\begin{pmatrix}1\\0\\0\end{pmatrix}$ &
  $\begin{pmatrix}0\\1\\0\end{pmatrix}$ &
  $\begin{pmatrix}0\\0\\1\end{pmatrix}$ &
  $\begin{pmatrix}-1\\0\\-1\end{pmatrix}$ &
  $\begin{pmatrix}-1\\-1\\0\end{pmatrix}$ &
  $\begin{pmatrix}1\\1\\1\end{pmatrix}$
 \end{tabular}
 \caption{Vertices of the three-dimensional reflexive polytope with PALP id $22$.}
 \label{t:rp22}
\end{table}
In order to confirm the existence of multiple $\rep{10}$ representations, we are therefore led to consider a nef partition
with non-trivial toric Mordell-Weil group. To be concrete, let us pick the following nef partition of the polytope given
in \autoref{t:rp22}:
\begin{equation} \label{eq:nef_fiber_rp22}
 \nabla_1 = \langle v_1 v_2 v_3 v_4 v_5 \rangle_{\textrm{conv}}\,, \qquad \nabla_2 = \langle v_0 \rangle_{\textrm{conv}}\,.
\end{equation}
Since $\nabla_2$ is one-dimensional, this nef partition is a \emph{projection}. In particular, this means that we
can directly solve the second equation, plug the result into the first equation and obtain the Weierstrass form
of a hypersurface equation. According to the conventions of \autoref{ss:nef_partitions}, this nef partition has the
unique id $(22, 0)$. Looking it up in our classification results,
we find that it has three sections, namely the divisors corresponding to the rays $v_1$, $v_2$, and $v_5$. Let us
divisor $V(z_5)$ as the neutral element of our elliptic curve. Then $\sigma_1 = V(z_0) - V(z_5)$ and
$\sigma_2 = V(z_2) - V(z_5)$ generate a $\mathbb{Z} \oplus \mathbb{Z}$ group.

Let us now write down the equations that define the complete intersection inside the three-dimensional toric
variety corresponding to the reflexive polytope of \autoref{t:rp22}. Keeping the coefficients general, the equations of the 
complete intersection defined by the nef partition \eqref{eq:nef_fiber_rp22} are
\begin{align}
 p_1 &= \tilde{a}_0 z_1^2 z_2^2 z_5^3 + \tilde{a}_1 z_1^2 z_2 z_3 z_5^2 + \tilde{a}_2 z_1 z_2^2 z_4 z_5^2
	+ \tilde{a}_3 z_1^2 z_3^2 z_5 + \tilde{a}_4 z_1 z_2 z_3 z_4 z_5 + \tilde{a}_5 z_2^2 z_4^2 z_5  \\
	& \quad + \tilde{a}_6 z_0 z_1 z_2 z_5^2 + \tilde{a}_7 z_1 z_3^2 z_4 + \tilde{a}_8 z_2 z_3 z_4^2
	+\tilde{a}_9 z_0 z_1 z_3 z_5 + \tilde{a}_{10} z_0 z_2 z_4 z_5 + \tilde{a}_{11} z_0 z_3 z_4
	+ \tilde{a}_{12} z_0^2 z_5 \nonumber \\
 p_2 &= \tilde{b}_0 z_1 z_2 z_5 + \tilde{b}_1 z_1 z_3 + \tilde{b}_2 z_2 z_4 + \tilde{b}_3 z_0\,.
\end{align}
Here one can see that this nef partition is indeed a projection: By solving $p_2=0$ for $z_0$ and inserting
the solution in $p_1$ the complete intersection is reduced to a hypersurface inside the toric variety corresponding
to the polytope obtained by projecting along $v_0$. However, this still suffices for our purposes.
Since it is the limited number of triangulations of the $SU(5)$ tops
for a codimension one hypersurface that constrains the $\rep{10}$ charges, we are still circumventing this constraint
here by considering triangulations of the higher-dimensional variety in which the elliptic curve has codimension two.

Next, we tune the $\tilde{a}_i$ and $\tilde{b}_i$ such as to enforce an $SU(5)$ singularity along the divisor
$e_0=0$ in the base manifold. Then
we resolve that singularity introducing exceptional divisors $e_i$, $i=1,\dots,4$ and find that the coefficients
$\tilde{a}_i$ and $\tilde{b}_i$ take the form
\begin{align} \label{eq:blowup_rp22}
 \tilde{a}_0 &= a_0 \cdot e_0^3 e_1 e_2^2 e_4^2 &
 \tilde{a}_1 &= a_1 \cdot e_0^2 e_1 e_2 e_4 &
 \tilde{a}_2 &= a_2 \cdot e_0^2 e_1 e_2^2 e_4 \nonumber \\
 \tilde{a}_3 &= a_3 \cdot e_0 e_1 &
 \tilde{a}_4 &= a_4 \cdot e_0 e_1 e_2 &
 \tilde{a}_5 &= a_5 \cdot e_0 e_1 e_2^2 \nonumber \\
 \tilde{a}_6 &= a_6 \cdot e_0 e_4 &
 \tilde{a}_7 &= a_7 \cdot e_0 e_1^2 e_2 e_3 &
 \tilde{a}_8 &= a_8 \cdot e_0 e_1^2 e_2^2 e_3 \nonumber \\
 \tilde{a}_9 &= a_9 \cdot e_0 e_1 e_3 e_4 &
 \tilde{a}_{10} &= a_{10} &
 \tilde{a}_{11} &= a_{11} \cdot e_1 e_3 \nonumber \\
 && \tilde{a}_{12} &= a_{12} \cdot e_0 e_1 e_3^2 e_4^2
\end{align}
and
\begin{align}
 \tilde{b}_0 &= b_0 \cdot e_0 e_2 e_4 &
 \tilde{b}_1 &= b_1 &
 \tilde{b}_2 &= b_2 \cdot e_2 &
 \tilde{b}_3 &= b_3 \cdot e_3 e_4\,.
\end{align}
Here $a_i$ and $b_i$ are polynomials in the base variables
that depend on $e_i$ only through the combination $w_0 \equiv e_0 e_1 e_2 e_3 e_4$.
\begin{table}[h]
 \centering
 \begin{tabular}{ccccc}
  $e_0$   & $e_1$ & $e_{2}$ & $e_{3}$ & $e_{4}$ \\
  \hline
  $\begin{pmatrix}0\\0\\0\\w_0\end{pmatrix}$ &
  $\begin{pmatrix}-1\\-1\\-1\\w_0\end{pmatrix}$ &
  $\begin{pmatrix}-1\\-1\\0\\w_0\end{pmatrix}$ &
  $\begin{pmatrix}0\\-1\\-1\\w_0\end{pmatrix}$ &
  $\begin{pmatrix}1\\0\\0\\w_0\end{pmatrix}$
 \end{tabular}
 \caption[$SU(5)$ top for the ambient space with PALP id $22$.]{Torically, the blowup of \eqref{eq:blowup_rp22} corresponds to introducing the top defined
 here, where $w_0$ is a ray of the fan of the base. The GUT brane will then be located on the divisor
 corresponding to $w_0$. Note that here we and in \eqref{eq:blowup_rp22} we are denoting the rays
 and the corresponding homogeneous variables by the same letters $e_i$.}
 \label{t:su5_top_rp22}
\end{table}
The toric data corresponding to this blowup are given in \autoref{t:su5_top_rp22}.

As a power series in $w_0$, the Weierstrass coefficients read
\begin{align}
 f &= -\frac{1}{48} \Big( a_{10}^4 \cdot b_1^4 + 
 4 \cdot a_{10}^{2} \cdot b_{1}^{2} \cdot c_1 \cdot w_0 
+ c_2 \cdot w_0^2  \Big) + \mathcal{O}(w_0^3) \\
 g &= \frac{1}{864} \Big( a_{10}^{6} \cdot b_{1}^{6} +
  6 \cdot a_{10}^{4} \cdot b_{1}^{4}  \cdot c_1 \cdot w_0 +
  3 b_{1}^{2} \cdot a_{10}^{2} \cdot c_3 \cdot w_0^2 + c_4 \cdot w_0^3 \Big) + \mathcal{O}(w_0^4) \,,
\end{align}
where the $c_i$ are irreducible polynomials in $a_i$ and $b_i$.
This implies that the discriminant $\Delta = 4 f^3 + 27 g^2$ takes the form
\begin{align}
 \Delta = \frac{1}{16}  \Big(
  a_{10}^{4} \cdot b_{1}^{4} \cdot  a_{11} \cdot b_{2} \cdot b_{3} 
  \cdot c_5 \cdot c_6 \cdot c_7 \cdot w_0^5
  + a_{10}^2 \cdot b_1^2 \cdot c_8 \cdot w_0^6 + c_9 \cdot w_0^7 \Big) + \mathcal{O}(w_0^8)
\end{align}
with
\begin{align}
 c_5 &= a_{10} a_{12} b_{1}^{2} -  a_{9} a_{10} b_{1} b_{3} + a_{6} a_{11} b_{1} b_{3} + a_{3} a_{10} b_{3}^{2} \\
 c_6 &= - a_{8} a_{10} b_{1}^{2} + a_{5} a_{11} b_{1}^{2} + a_{7} a_{10}
b_{1} b_{2} -  a_{4} a_{11} b_{1} b_{2} + a_{3} a_{11} b_{2}^{2} \\
c_7 &= a_{3} a_{10}^{2} b_{0}^{2} + a_{4} a_{6} a_{10} b_{0} b_{1} -  a_{1}
a_{10}^{2} b_{0} b_{1} + a_{5} a_{6}^{2} b_{1}^{2} -  a_{2} a_{6} a_{10}
b_{1}^{2} + a_{0} a_{10}^{2} b_{1}^{2} \nonumber \\
& \quad - 2 a_{3} a_{6} a_{10} b_{0}
b_{2} -  a_{4} a_{6}^{2} b_{1} b_{2} + a_{1} a_{6} a_{10} b_{1} b_{2} +
a_{3} a_{6}^{2} b_{2}^{2}\,.
\end{align}
From the vanishing orders of the $f$, $g$ and $\Delta$ we observe
that there are seven distinct matter curves and list them in \autoref{t:rp22_curves}.
\begin{table}[h]
 \centering
 \begin{tabular}{cccc}
  Name & Equation & Singularity type & $SU(5)$ representation \\
  \hline
  \hline
  $T_1$ & $a_{10} \cap w_0$ & $SO(10)$ & $\rep{10}$  \\
  $T_2$ & $b_1 \cap w_0$ & $SO(10)$ & $\rep{10}$  \\
  \hline
  $F_1$ & $a_{11} \cap w_0$ & $SU(7)$ & $\rep{5}$  \\
  $F_2$ & $b_{2} \cap w_0$ & $SU(7)$ & $\rep{5}$  \\
  $F_3$ & $c_{5} \cap w_0$ & $SU(7)$ & $\rep{5}$  \\
  $F_4$ & $c_{6} \cap w_0$ & $SU(7)$ & $\rep{5}$  \\
  $F_5$ & $c_{7} \cap w_0$ & $SU(7)$ & $\rep{5}$  \\
  $F_6$ & $b_{3} \cap w_0$ & $SU(7)$ & $\rep{5}$   \\
 \end{tabular}
 \caption{The matter curves for the top of \autoref{t:su5_top_rp22}.}
  \label{t:rp22_curves}
\end{table}
\begin{table}[!ht]
 \centering
 \begin{tabular}{ccc}
  Singularity type & Coupling & Multiplicity\\
  \hline
  \hline
  $SU(7)$ & $\rep{5}_{(4, 3)} \times \overline{\rep{5}}_{(1, 2)}$ & $54$ \\
  $SU(7)$ & $\rep{5}_{(-1,3)} \times \overline{\rep{5}}_{(1, 2)}$ & $39$ \\
  $SU(7)$ & $\rep{5}_{(-1,3)} \times \overline{\rep{5}}_{(-4, -3)}$ & $36$ \\
  $SU(7)$ & $\rep{5}_{(-6, -7)} \times \overline{\rep{5}}_{(1, 2)}$ & $27$ \\
  $SU(7)$ & $\rep{5}_{(-6, -7)} \times \overline{\rep{5}}_{(-4, -3)}$ & $12$ \\
  $SU(7)$ & $\rep{5}_{(-6, -7)} \times \overline{\rep{5}}_{(1,-3)}$ & $9$ \\
  $SU(7)$ & $\rep{5}_{(-6, -2)} \times \overline{\rep{5}}_{(1, 2)}$ & $9$ \\
  $SU(7)$ & $\rep{5}_{(-6, -2)} \times \overline{\rep{5}}_{(-4, -3)}$ & $6$ \\
  $SU(7)$ & $F\rep{5}_{(-6, -2)} \times \overline{\rep{5}}_{(1,-3)}$ & $6$ \\
  $SU(7)$ & $\rep{5}_{(-6, -7)} \times \overline{\rep{5}}_{(6, 2)}$ & $3$ \\
  \hline
  $SO(12)$ & $\overline{\rep{10}}_{(-3,-1)} \times \rep{5}_{(4, 3)} \times \rep{5}_{(-1, -2)}$ & $15$ \\
  $SO(12)$ & $\overline{\rep{10}}_{(2, 4)} \times \rep{5}_{(-1, -2)} \times \rep{5}_{(-1, -2)}$ & $3$ \\
  $SO(12)$ & $\overline{\rep{10}}_{(2, 4)} \times \rep{5}_{(-6, -2)} \times \rep{5}_{(4, 3)}$ & $3$ \\
  \hline
  $E_6$ & $\rep{10}_{(3,1)} \times \rep{10}_{(3,1)} \times \rep{5}_{(-6, -2)}$ & $3$ \\
  $E_6$ & $\rep{10}_{(3,1)} \times \rep{10}_{(-2, -4)} \times \rep{5}_{(-1,3)}$ & $3$ \\
 \end{tabular}
 \caption[Yukawa couplings for $SU(5)$ top with fiber with PALP id $22$.]{All couplings involving multiple non-Abelian 
 matter representations in the example of \autoref{eq:rp22_fibration}.
 Note that there are additional non-minimal singularities that do not list here.}
 \label{t:rp22_couplings}
\end{table}

While the appearance of two different $\rep{10}$ curves and six distinct $\rep{5}$ curves is promising, it is crucial to
check which of these curves are actually realized in a generic fibration of this top over a base manifold.
Next, we therefore fiber this space over a $\mathbb{P}^3$. Doing so can be achieved by embedding the rays of
\autoref{t:rp22} into $\mathbb{Z}^6$ according to
\begin{align}
 v_i \mapsto u_i \equiv (v_i, 0, 0, 0), \qquad i = 1, \ldots, 5\,,
\end{align}
adding the blowup rays from \autoref{t:su5_top_rp22} with $w_0 = (1, 0, 0)$ and adding the remaining $3$ base rays:
\begin{align} \label{eq:rp22_fibration}
  u_7 = \left(0,0,0,-1,-1,-1\right)\,, \quad
  u_8 = \left(n_1, n_2, n_3, 0, 1, 0\right)\,, \quad
  u_9 = \left(0, 0, 0, 0, 0, 1\right)\,.
\end{align}
Here the $n_i$ are integers encoding the fibration of the fiber over the base. More specifically,
the $n_i$ determine which line bundles the fiber coordinates are sections of. For our purposes,
we choose $(n_1, n_2, n_3) = (-1, 0, 0)$.
After using PALP to compute all nef partitions of the resulting polytope, we pick the one with
\begin{align}
\nabla_1 = \langle u_1, u_2, u_3, u_4, u_5, u_6, u_7, u_8, e_0, e_{1}, e_{2} \rangle_{\textrm{conv}}\,,
\qquad \nabla_2 = \langle u_0, e_3, e_4 \rangle_{\textrm{conv}}\,.
\end{align}
It has Hodge numbers $h^{1,1} = 8$, $h^{2,1} = 0$, and $h^{3,1} = 141$. For this specific choice of fibration,
both $b_0$ and $b_3$ are constants.
Consequently, the curve $F_6$ is not realized. However, all other curves \emph{exist} and in particular,
there are two different antisymmetric representations. 
Using the Chern-Simons matching as in \cite{Grimm:2011fx, Cvetic:2012xn, Grimm:2013oga}, we find that the realized curves
have the following charges under the two $U(1)$s:
\begin{align}
 T_1:\ \rep{10}_{(3,1)}\,, \qquad  T_2:\ \rep{10}_{(-2, -4)}
\end{align}
\begin{align}
 F_1:\ \rep{5}_{(-6, -7)}\,, \quad F_2:\ \rep{5}_{(-6, -2)}\,, \quad F_3: \ \rep{5}_{(-1,3)}\,, 
 \quad F_4: \ \rep{5}_{(4, 3)}\,, \quad F_5:\ \rep{5}_{(-1, -2)}
\end{align}
We also find the following singlet states:
\begin{align}
 \rep{1}_{(5,0)}\,, \quad  \rep{1}_{(0,5)}\,,\quad  \rep{1}_{(5,5)}\,,\quad \rep{1}_{(5,10)}\,,
 \quad \rep{1}_{(10,5)}\,,\quad \rep{1}_{(10, 10)}\,.
\end{align}
Finally, we compute the Yukawa couplings for the given example and find the ones listed in
\autoref{t:rp22_couplings}.

In summary, we have managed to construct a fully explicit F-theory model with gauge group $SU(5) \times U(1)^2$,
in which the \emph{torically realized} $SU(5)$
singularity gives rise to a gauge theory with two different $\rep{10}$ representations. Clearly the example
studied here is not intended to be used as a full-fledged GUT model. In more realistic models
several issues would need to be addressed, such as the fact that there exist non-minimal singularities
at points in the base manifold whose resolution leads to a non-flat fibration. Furthermore, the topology
of the GUT divisor is too simple in order to allow hypercharge flux with the desired properties.
In principle, both these points can be addressed by choosing the fibration more carefully than we did following
equation \eqref{eq:rp22_fibration}.

\subsection
[\texorpdfstring{$SU(5) \times U(1)^2$}{SU(5)xU(1)xU(1)} and a Discrete Symmetry]
{\boldmath $SU(5) \times U(1)^2$ and a Discrete Symmetry} 
\label{ss:example_p3}

The second example we consider is a nef partition of the polytope with the least integral points,
that is the one corresponding to $\mathbb{P}^3$. Its polytope is of course well-known, but for completeness
we list it in \autoref{t:p3}.
\begin{table}[h]
 \centering
 \begin{tabular}{cccccc}
  $v_0$ & $v_1$ & $v_2$ & $v_3$ &\\
  \hline
  $\begin{pmatrix}-1\\-1\\-1\end{pmatrix}$ &
  $\begin{pmatrix}0\\0\\1\end{pmatrix}$ &
  $\begin{pmatrix}0\\1\\0\end{pmatrix}$ &
  $\begin{pmatrix}1\\0\\0\end{pmatrix}$
 \end{tabular}
 \caption[Vertices of the three-dimensional reflexive polytope with PALP id $0$.]{Vertices
 of the reflexive polytope corresponding to $\mathbb{P}^3$. Since it has the least integral points of
 all reflexive polytopes in three dimensions, it has PALP id $0$.}
 \label{t:p3}
\end{table}
All toric divisors $V(z_i)$ inside $\mathbb{P}^3$ lie in the same homology class and therefore it can only have two nef partitions:
The one corresponding to a partition of $3+1$ vertices and the nef partition corresponding to a partition of $2+2$ vertices.
The first is again a projection and to have some variety, we therefore focus on the latter. That is, we take
our nef partition to be
\begin{align} \label{eq:nef_rp0}
 \nabla_1 = \langle v_0, v_3\rangle_{\textrm{conv}}\,, \qquad  \nabla_2 = \langle v_1, v_2\rangle_{\textrm{conv}}\,.
\end{align}
This implies automatically that all toric divisors intersect a generic complete intersection of this type in four points:
\begin{align}
 V(z_i) \cap E = \int_E [V(z_i)] = \int_{\mathbb{P}^3} [2H] \cdot [2H] \cdot [H] = 4\,.
\end{align}
A generic fibration with this fiber will therefore not have a section. As noted in the introduction,
F-theory models without section have recently received quite some attention, see \cite{Braun:2014oya,
Morrison:2014era,Anderson:2014yva,Klevers:2014bqa,Garcia-Etxebarria:2014qua, Mayrhofer:2014haa}.
However, in these models the Calabi-Yau manifolds always had $2$- or $3$-sections leading to $\mathbb{Z}_2$
or $\mathbb{Z}_3$ discrete gauge symmetries, respectively. As the biquadric in $\mathbb{P}^3$ has a $4$-section,
we expect to find a discrete $\mathbb{Z}_4$ gauge group. In the following we will try to collect
some further evidence for this.

To do, let us take the same approach as with the previous example and write down the defining
equations of the complete intersection. They read
\begin{align} \label{eq:rp0}
 p_1 &= \tilde{a}_0 z_{0}^{2} + \tilde{a}_1 z_{0} z_{1} + \tilde{a}_2 z_{1}^{2}
+ \tilde{a}_3 z_{0} z_{2} + \tilde{a}_4 z_{1} z_{2} + \tilde{a}_5
z_{2}^{2} + \tilde{a}_6 z_{0} z_{3} + \tilde{a}_7 z_{1} z_{3} +
\tilde{a}_8 z_{2} z_{3} + \tilde{a}_9 z_{3}^{2} \nonumber \\
 p_2 &= \tilde{b}_0 z_{0}^{2} + \tilde{b}_1 z_{0} z_{1} + \tilde{b}_2 z_{1}^{2}
+ \tilde{b}_3 z_{0} z_{2} + \tilde{b}_4 z_{1} z_{2} + \tilde{b}_5
z_{2}^{2} + \tilde{b}_6 z_{0} z_{3} + \tilde{b}_7 z_{1} z_{3} +
\tilde{b}_8 z_{2} z_{3} + \tilde{b}_9 z_{3}^{2}\,.
\end{align}
Note that such biquadrics have been studied before in \cite{Esole:2011cn} and, with the restriction to the triple
blowup of $\mathbb{P}^3$, in \cite{Cvetic:2013qsa}. Since this nef partition is not a projection, one cannot
bring this complete intersection into Weierstrass form by solving one of the equations for one
variable and substituting the result into the other equation.

\begin{table}[h]
 \centering
 \begin{tabular}{ccccc}
  $e_0$   & $e_1$ & $e_{2}$ & $e_{3}$ & $e_{4}$ \\
  \hline
  $\begin{pmatrix}0\\0\\0\\w_0\end{pmatrix}$ &
  $\begin{pmatrix}-1\\-1\\-1\\w_0\end{pmatrix}$ &
  $\begin{pmatrix}-1\\-1\\0\\w_0\end{pmatrix}$ &
  $\begin{pmatrix}0\\-1\\0\\w_0\end{pmatrix}$ &
  $\begin{pmatrix}0\\-1\\-1\\w_0\end{pmatrix}$
 \end{tabular}
 \caption[$SU(5)$ top for the biquadric in $\mathbb{P}^3$.]{As before, the blowup of equations \eqref{eq:blowup_rp0_1} and \eqref{eq:blowup_rp0_2}
 corresponds to introducing the top defined here,
 where $w_0$ is a ray of the fan of the base. The GUT brane will then be located on the divisor
 corresponding to $w_0$. We again denote rays and corresponding homogeneous variables by the same letters.}
 \label{t:su5_top_rp0}
\end{table}
Next, we tune the $\tilde{a}_i$ and $\tilde{b}_i$ such as to enforce an $SU(5)$ singularity along the divisor
$e_0=0$ in the base manifold. Then we resolve this singularity by introducing exceptional divisors $e_i$, $i=1,\dots,4$
as specified torically in terms of the top of \autoref{t:su5_top_rp0}.
We find that the coefficients $\tilde{a}_i$ and $\tilde{b}_i$ take the form
\begin{align} \label{eq:blowup_rp0_1}
 \tilde{a}_0 &= a_0 \cdot e_{1}^{2} e_{2}^{2} e_{3} e_{4} &
 \tilde{a}_1 &= a_1 \cdot e_{1} e_{2}^{2} e_{3}&
 \tilde{a}_2 &= a_2 \cdot e_{0} e_{1} e_{2}^{3} e_{3}^{2}  \nonumber \\
 \tilde{a}_3 &= a_3 \cdot e_1 e_2 &
 \tilde{a}_4 &= a_4 \cdot e_{0} e_{1} e_{2}^{2} e_{3} &
 \tilde{a}_5 &= a_5 \cdot e_0 e_1 e_2 \nonumber \\
 \tilde{a}_6 &= a_6 \cdot e_1 e_2 e_3 e_4 &
 \tilde{a}_7 &= a_7 \cdot e_2 e_3 &
 \tilde{a}_8 &= a_8  \nonumber \\ &&
 \tilde{a}_9 &= a_9 \cdot  e_3 e_4 &
\end{align}
and
\begin{align} \label{eq:blowup_rp0_2}
 \tilde{b}_0 &= b_0 \cdot e_1 e_4 &
 \tilde{b}_1 &= b_1 &
 \tilde{b}_2 &= b_2 \cdot e_0 e_2 e_3 \nonumber \\
 \tilde{b}_3 &= b_3 \cdot e_0 e_1 e_4 &
 \tilde{b}_4 &= b_4 \cdot e_0 &
 \tilde{b}_5 &= b_5 \cdot e_0^2 e_1 e_4 \nonumber \\
 \tilde{b}_6 &= b_6 \cdot e_0 e_1 e_3 e_4^2 &
 \tilde{b}_7 &= b_7 \cdot e_0 e_3 e_4 &
 \tilde{b}_8 &= b_8 \cdot e_0^2 e_1 e_3 e_4^2 \nonumber \\
 && \tilde{b}_9 &= b_9 \cdot e_0^2 e_1 e_3^2 e_4^3 \,.
\end{align}
Here $a_i$ and $b_i$ are polynomials in the base variables
that depend on $e_i$ only through the combination $w_0 \equiv e_0 e_1 e_2 e_3 e_4$.
As a power series in $w_0$, the Weierstrass coefficients read
\begin{align}
 f &= -\frac{1}{768} \Big( a_{8}^4 \cdot b_1^4 + 
 2 \cdot a_{8}^{2} \cdot b_{1}^{2} \cdot c_1 \cdot w_0 
+ c_2 \cdot w_0^2  \Big) + \mathcal{O}(w_0^3) \\
 g &=  \frac{1}{55296} \Big( a_8^6 \cdot b_1^6 
 - 3 \cdot a_8^4 \cdot b_1^4 \cdot c_1 \cdot w_0
 +  a_8^2 \cdot b_1^2 \cdot c_3 \cdot w_0^2
 + c_4 \cdot w_0^3 \Big) + \mathcal{O}(w_0^4)\,,
\end{align}
where the $c_i$ are irreducible polynomials in $a_i$ and $b_i$.
Then the discriminant is
\begin{align}
 \Delta = \frac{1}{2^{16}}  \Big( a_8^4 \cdot b_1^4 \cdot c_5 \cdot c_6 \cdot c_7 \cdot c_8 \cdot w_0^5
 + a_8^2 \cdot b_1^2 \cdot c_9 \cdot v_0^6 + c_{10} \cdot w_0^7
 \Big) + \mathcal{O}(w_0^8)
\end{align}
with
\begin{align}
 c_5 &= -b_1 b_3 b_4 + b_0 b_4^2 + b_1^2 b_5 \\
 c_6 &= a_3 a_7 a_8 b_0 - a_1 a_8^2 b_0 - a_3 a_6 a_8 b_1 + a_0 a_8^2 b_1 + a_3^2 a_9 b_1 \\
c_7 &= -a_5 a_7^2 b_1 + a_4 a_7 a_8 b_1 - a_2 a_8^2 b_1 - a_3 a_7 a_8 b_2
+ a_1 a_8^2 b_2 +  a_3 a_7^2 b_4 - a_1 a_7 a_8 b_4 \\
 c_8 &= -a_9^2 b_1 b_3 b_4 + a_9^2 b_0 b_4^2 + a_9^2 b_1^2 b_5 + a_8 a_9 b_1 b_4 b_6 + 
 a_8 a_9 b_1 b_3 b_7 - 2 a_8 a_9 b_0 b_4 b_7 \nonumber \\
 & \quad\, - a_8^2 b_1 b_6 b_7 + a_8^2 b_0 b_7^2 - 
 a_8 a_9 b_1^2 b_8 + a_8^2 b_1^2 b_9
 \,.
\end{align}
We observe that there are six distinct matter curves and list them in \autoref{t:p3_curves}.
This by itself is another piece of evidence that there exists in fact an order $4$ discrete symmetry.
Arguing along the lines of \cite{Garcia-Etxebarria:2014qua, Mayrhofer:2014haa}, it is this
symmetry that helps to distinguish the four $\rep{5}$ representations that would otherwise
have identical quantum numbers in the low-energy effective action.
\begin{table}[h]
 \centering
 \begin{tabular}{cccc}
  Name & Equation & Singularity type & $SU(5)$ representation \\
  \hline
  \hline
  $T_1$ & $a_{8} \cap w_0$ & $SO(10)$ & $\rep{10}$  \\
  $T_2$ & $b_1 \cap w_0$ & $SO(10)$ & $\rep{10}$ \\
  \hline
  $F_1$ & $c_5 \cap w_0$ & $SU(7)$ & $\rep{5}$  \\
  $F_2$ & $c_6 \cap w_0$ & $SU(7)$ & $\rep{5}$  \\
  $F_3$ & $c_{7} \cap w_0$ & $SU(7)$ & $\rep{5}$   \\
  $F_4$ & $c_{8} \cap w_0$ & $SU(7)$ & $\rep{5}$   \\
 \end{tabular}
 \caption{The matter curves in the example with the elliptic fiber embedded
	  as a biquadric in $\mathbb{P}^3$.}
  \label{t:p3_curves}
\end{table}
\begin{table}[h]
 \centering
 \begin{tabular}{ccc}
  Singularity type & Coupling & Multiplicity \\
  \hline
  \hline
  $SU(7)$ & $F_1 \times F_2$ & $30$ \\
  $SU(7)$ & $F_1 \times F_3$ & $42$ \\
  $SU(7)$ & $F_1 \times F_4$ & $36$ \\
  $SU(7)$ & $F_2 \times F_3$ & $33$ \\
  $SU(7)$ & $F_2 \times F_4$ & $40$ \\
  $SU(7)$ & $F_3 \times F_4$ & $56$ \\
  \hline
  $SO(12)$ & $T_1 \times F_1 \times F_4$ & $6$ \\
  $SO(12)$ & $T_1 \times F_2 \times F_2$ & $1$ \\
  $SO(12)$ & $T_1 \times F_3 \times F_3$ & $2$ \\
  $SO(12)$ & $T_2 \times F_1 \times F_1$ & $6$ \\
  $SO(12)$ & $T_2 \times F_2 \times F_3$ & $9$ \\
  $SO(12)$ & $T_2 \times F_4 \times F_4$ & $9$ \\
  \hline
  $E_6$ & $T_1 \times T_1 \times F_3$ & $3$ \\
  $E_6$ & $T_1 \times T_2 \times F_2$ & $3$ \\
  $E_6$ & $T_2 \times T_2 \times F_3$ & $12$
 \end{tabular}
\caption[Yukawa couplings for $SU(5)$ top with fiber $\mathbb{P}^3$.]{All
couplings involving multiple non-Abelian matter representations in the example 
  with the elliptic fiber embedded in $\mathbb{P}^3$.
 Note that there are additional non-minimal singularities that do not list here.}
 \label{t:rp0_couplings}
\end{table}

As before, we can make this more concrete by constructing an explicit example. To do so,
we use the same embedding into $\mathbb{Z}^6$ as in equation \eqref{eq:rp22_fibration},
but this time we set $(n_1, n_2, n_3) = (0,0,1)$ and denote the rays obtained by
embedding the base divisors $w_i$, $i=1,2,3$ by $u_5$, $u_6$, and $u_7$. 
The resulting six-dimensional lattice
polytope has $33$ nef partitions. Of these, let us pick the nef partition
\begin{align} \label{eq:nef_fibration_rp0}
 \nabla_1 = \langle u_0, u_3, u_5, e_1, e_2, e_3, e_4 \rangle_{\textrm{conv}}\,, \qquad
 \nabla_2 = \langle u_1, u_2, e_0, u_6, u_7 \rangle_{\textrm{conv}}\,,
\end{align}
which has the Hodge numbers $h^{1,1} = 6$, $h^{2,1} = 0$, and $h^{3,1} = 110$.
For this explicit example, we find that all the curves listed in \autoref{t:p3_curves}
are in fact realized geometrically. In \autoref{t:rp0_couplings} we furthermore list
the Yukawa points involving multiple non-Abelian representations. Since Yukawa couplings must
be invariant under gauge symmetries, the couplings that do not involve singlets allow us to determine the $\mathbb{Z}_4$
charges of the six matter curves. Let us denote the neutral element of $\mathbb{Z}_4$ by
$0$ and call the generator $e$. Then we have that the two couplings involving only $T_1$ and $F_3$
imply
\begin{align}
 2 \cdot Q_{\mathbb{Z}_4}(T_1) + Q_{\mathbb{Z}_4}(F_3) = 0\,,
 \qquad 2 \cdot Q_{\mathbb{Z}_4}(F_3) = T_1
\end{align}
which immediately leads to
\begin{align}
 Q_{\mathbb{Z}_4}(T_1) = Q_{\mathbb{Z}_4}(F_3) = 0\,.
\end{align}
The remaining couplings then imply that
\begin{align}
 Q_{\mathbb{Z}_4}(F_2) = Q_{\mathbb{Z}_4}(T_2) = 2 e\,.
\end{align}
Last but not least, we have $Q_{\mathbb{Z}_4}(F_{1 / 4}) \in \{e, 3e\}$. However, $e$ and $3e$ are
the only order $4$ elements of $\mathbb{Z}_4$ and we could just as well take $e' = 3e$ as the generator
of $\mathbb{Z}_4$. As a consequence, one can simply choose that
\begin{align}
 Q_{\mathbb{Z}_4}(F_1) = e\,, \qquad Q_{\mathbb{Z}_4}(F_4) = 3 e\,.
\end{align}
With these charge assignments one finds that singlets with all allowed $\mathbb{Z}_4$ charges must be present
in order to make all the couplings of \autoref{t:rp0_couplings} invariant.

Put in a nutshell, we find that one can easily realize F-theory models with a non-Abelian gauge group
accompanied solely by an additional discrete symmetry of order $4$. A convenient way of doing so
proceeds by embedding the elliptic fiber as a biquadric inside $\mathbb{P}^3$. There
are numerous ways of extending the treatment here, such as connecting this model
to others in terms of Higgsings and conifold transitions in the circle-compactified
theories.

\subsection
[Example with Mordell-Weil Torsion \texorpdfstring{$\mathbb{Z}_4$}{Z4}]
{Example with Mordell-Weil Torsion \boldmath$\mathbb{Z}_4$}

As a final example, let us take a quick look at a model with Mordell-Weil torsion $\mathbb{Z}_4$. This torsion group
does not exist generically for codimension one elliptic fibers \cite{Braun:2013nqa, Mayrhofer:2014opa,Klevers:2014bqa}
and even in codimension two there is only a single example as can be seen from \autoref{t:mw_groups}.

Mordell-Weil torsion was studied extensively in \cite{Mayrhofer:2014opa} and it was found that it
impacts the global structure of the non-Abelian gauge group. Given a singularity of type $A_{n-1}$,
the universal covering group is $SU(n)$, which, without Mordell-Weil torsion, constitutes the gauge group of the F-theory model.
In the presence of a non-trivial Mordell-Weil torsion group $\mathbb{Z}_k$ this changes: The non-Abelian gauge group becomes $SU(n) / \mathbb{Z}_k$.
By construction the universal covering group has a trivial first fundamental group, and therefore the effect of non-trivial Mordell-Weil
torsion is that the non-Abelian gauge group of the low-energy effective theory is no longer simply connected:
\begin{align}
 \pi_1(SU(n) / \mathbb{Z}_k) = \mathbb{Z}_k\,.
\end{align}
In the examples studied in \cite{Mayrhofer:2014opa} Mordell-Weil torsion groups $\mathbb{Z}_2$ and $\mathbb{Z}_3$ always
came accompanied by gauge groups of type $SU(2n)$ and $SU(3n)$, respectively. Since $SU(n)$ has a $\mathbb{Z}_n$
center generated by the identity matrix times $e^{\frac{2 \pi i}{n}}$, one can mod out $\mathbb{Z}_k$ by eliminating the center
(or a subgroup thereof) of $SU(k \cdot n)$. 

\begin{table}[h]
 \centering
 \begin{tabular}{cccccccc}
  $v_0$ & $v_1$ & $v_2$ & $v_3$ & $v_4$ & $v_5$ & $v_6$ & $v_7$ \\
  \hline
  $\begin{pmatrix} 1\\0\\0\end{pmatrix}$ &
  $\begin{pmatrix} 0\\1\\0\end{pmatrix}$ &
  $\begin{pmatrix} 1\\-1\\0\end{pmatrix}$ &
  $\begin{pmatrix} -1\\0\\0\end{pmatrix}$ &
  $\begin{pmatrix} 0\\1\\2\end{pmatrix}$ &
  $\begin{pmatrix} -1\\0\\-2\end{pmatrix}$ &
  $\begin{pmatrix} -1\\-2\\-2\end{pmatrix}$ &
  $\begin{pmatrix} 2\\1\\2\end{pmatrix}$
 \end{tabular}
 \caption{Vertices of the three-dimensional reflexive polytope with PALP id $3415$.}
 \label{t:rp3415}
\end{table}

The corresponding reflexive polytope has PALP id $3415$ and we list its defining data in \autoref{t:rp3415}.
It has a single nef partition, namely
\begin{align} \label{eq:nef_3415}
 \nabla_1 = \langle v_0, v_3, v_5, v_6 \rangle_{\textrm{conv}}\,, \qquad
 \nabla_2 = \langle v_1, v_2, v_4, v_7 \rangle_{\textrm{conv}}\,.
\end{align}
\begin{table}
 \centering
 \begin{tabular}{ccccccccccc}
  $v_8$ & $v_9$ & $v_{10}$ & $v_{11}$ & $v_{12}$ & $v_{13}$ &
  $v_{14}$ & $v_{15}$ & $v_{16}$ & $v_{17}$ & $v_{18}$ \\
  \hline
  $\begin{pmatrix}0\\ -1\\ -1\end{pmatrix}$ &
  $\begin{pmatrix}1\\ 0\\ 1 \end{pmatrix}$ &
  $\begin{pmatrix}-1\\ -1\\ -1 \end{pmatrix}$ &
  $\begin{pmatrix}0\\ 0\\ 1\end{pmatrix}$ &
  $\begin{pmatrix}-1\\ -1\\ -2\end{pmatrix}$ &
  $\begin{pmatrix}0\\ 0\\ 0\end{pmatrix}$ &
  $\begin{pmatrix}1\\ 1\\ 2\end{pmatrix}$ &
  $\begin{pmatrix}0\\ 0\\ -1\end{pmatrix}$ &
  $\begin{pmatrix}1\\ 1\\ 1\end{pmatrix}$ &
  $\begin{pmatrix}-1\\ 0\\ -1\end{pmatrix}$ &
  $\begin{pmatrix}0\\ 1\\ 1\end{pmatrix}$
 \end{tabular}
 \caption[Further integral points of polytope with PALP id $3415$.]{Integral points of the reflexive polytope with PALP id $3415$ that are neither vertices nor the origin.
 In order to fully resolve every fibration of the nef partition \eqref{eq:nef_3415} one must use all of these points
 as rays of the toric fan.}
 \label{t:rp3415_points}
\end{table}
In order to write down the most general complete intersection corresponding to this nef partition, we must
use every integral point of the polytope defined in \autoref{t:rp3415} apart from the origin. The additional
eleven points are listed in \autoref{t:rp3415_points}.

After resolution, the complete intersection defined by \eqref{eq:nef_3415} is defined by the following two polynomials:
\begin{align}
 p_1 &= a_0 z_0 z_3 z_5 z_6 z_8 z_{10} z_{12} z_{15} z_{17}  + a_1 z_0^2 z_7^2 z_8 z_9 z_{14} z_{15} z_{16} 
 + a_2 z_3^2 z_4^2 z_{10} z_{11} z_{14} z_{17} z_{18} \nonumber \\
 p_2 &= b_0 z_1^2 z_5^2 z_{12} z_{15} z_{16} z_{17} z_{18} + b_1 z_2^2 z_6^2 z_8 z_9 z_{10} z_{11} z_{12}
 + b_2 z_1 z_2 z_4 z_7 z_ 9 z_{11} z_{14} z_{16} z_{18}\,.
\end{align}
This time we are not interested in engineering additional singularities, but rather in confirming that models with
this fiber contain the $SU(4)$ gauge factors that we expect to exist. To this end we compute the discriminant of the
elliptic curve and find
\begin{align}
 f &= -\frac{1}{48} \cdot \left( 16 a_1^2 a_2^2 b_0^2 b_1^2 - 16 a_0^2 a_1 a_2 b_0 b_1 b_2^2 + a_0^4 b_2^4\right) \\
 g &= \frac{1}{864} \cdot \left( 8 a_1 a_2 b_0 b_1 - a_0^2 b_2^2\right) \cdot 
 \left(8 a_1^2 a_2^2 b_0^2 b_1^2 + 16 a_0^2 a_1 a_2 b_0 b_1 b_2^2 - a_0^4 b_2^4 \right) \\
 \Delta &= -\frac{1}{16} \cdot a_0^2 \cdot b_2^2 \cdot a_1^4 \cdot a_2^4  \cdot b_0^4 \cdot b_1^4
 \cdot \left(-16 a_1 a_2 b_0 b_1 + a_0^2 b_2^2\right)\,.
\end{align}
From the vanishing orders we see that there are two $I_2$ and four $I_4$ singularities. Since
\begin{align}
 \frac{9g}{2f} \Big\rvert_{a_1=0} = \frac{9g}{2f} \Big\rvert_{a_2=0} = \frac{9g}{2f} \Big\rvert_{b_1=0} = \frac{9g}{2f} \Big\rvert_{b_2=0}
 = -\frac{1}{4} a_0^2 b_3^2
\end{align}
the $I_4$ singularities are of split type (see \cite{Grassi:2011hq} or \autoref{a:groups})
and we therefore see that there is indeed a non-toric $SU(2)^2 \times SU(4)^4 / \mathbb{Z}_4$ gauge group. One can mod out
the $\mathbb{Z}_4$ torsion by identifying it with the diagonal subgroup of the center $\mathbb{Z}_4^{\oplus 4}$ of the
$SU(4)$ gauge group part.

It is interesting to see that up to a lattice isomorphism the reflexive polytope $\nabla^\circ$ associated to the nef partition \eqref{eq:nef_3415}
is precisely the polytope with PALP id $0$. Under the same lattice isomorphism, the $\Delta_i$ of \eqref{eq:nef_3415}
are mapped to the $\nabla_i$ of \eqref{eq:nef_rp0} and we therefore see that the fiber considered in this subsection
is mirror-dual to the fiber of \autoref{ss:example_p3}. In particular, it appears that under this duality
the discrete gauge group part is mapped to the torsion part of the Mordell-Weil group and vice versa.
The same behavior was observed in \cite{Klevers:2014bqa} for hypersurface fibers and it is intriguing to speculate
about a possible physical reason underlying this observation.

Finally, let us note that it would be interesting to study explicit realizations of such fibrations. While this is possible
in principle, the large number of involved points might make it technically challenging to find a triangulation that gives
rise to an appropriate toric fan of the ambient variety. In the recent work \cite{Long:2014fba} it was used that the relevant
triangulations are star triangulations with respect to the origin in order to speed up the calculation. It would exciting
to incorporate such an algorithm in the Sage software package and
apply it to these spaces.


 \section{Conclusions} \label{s:conclusions}

In this paper we proposed a new algorithm to bring a large class of elliptic curves
as well as the Jacobians of genus-one curves into Weierstrass form. The essential step of this algorithm
is to obtain an appropriate line bundle whose sections can be used as coordinates for an embedding
into either $\mathbb{P}_{231}$, $\mathbb{P}_{112}$, $\mathbb{P}^2$, or $\mathbb{P}^3$.
While it is not always possible to identify such a line bundle, the class for which this can be achieved
is much larger than the class of models that one has so far been able to bring into Weierstrass form.
To illustrate this fact, we computed the Weierstrass forms of all nef partitions
of three-dimensional reflexive polytopes that do not correspond to product spaces,
which allowed us to compute the toric Mordell-Weil group of all $3134$ complete intersection
curves of codimension two. Compared to the analogous analysis for hypersurfaces \cite{Braun:2013nqa},
we find additional groups, such as a free Abelian group of rank four or the pure torsion group $\mathbb{Z}_4$.
Additionally, we computed the non-toric non-Abelian gauge groups and again found a considerably larger
variety of than those that were encountered in \cite{Klevers:2014bqa} for hypersurface fibers.

In \autoref{s:examples} we proceeded by selecting three particular examples that exhibit features
that are ruled out for hypersurface fibers. These are torically realized $SU(5)$ models whose
antisymmetric representations have different charges under the additional
Abelian factors, models with a discrete $\mathbb{Z}_4$ symmetry and, finally,
F-theory models with a $\mathbb{Z}_4$ torsion factor.
For the first two types we give an explicit toric realization with non-Abelian gauge group $SU(5)$
and determine the matter curves that are present as well as the Yukawa couplings that the non-Abelian
representations are involved in.

There are numerous exciting ways in which this work could be extended in the future. On the one hand,
there are systematic questions that one could address, such as a classification of higher-dimensional
tops encoding the toric gauge groups or the construction of all fibrations with a given top.
For hypersurfaces these questions have already been answered in \cite{Bouchard:2003bu} and
\cite{Braun:2013nqa}, respectively, but it would be interesting to see how these result generalize
to higher codimensions.
On the other hand, one could use the methods developed here in order to construct explicit
scenarios for studying new physical effects. \autoref{s:examples} dealt with some potentially
interesting set-ups, but naturally there exist many more. Viewed more generally, one could hope
that access to a large number of fiber types might allow one to make observations about the landscape of
F-theory models \cite{Morrison:2012np,Morrison:2012js,Grimm:2012yq,Martini:2014iza}.
In \cite{Klevers:2014bqa} such observations were made based on the results for the
$16$ hypersurface fibers and, for instance, a network of Higgsing transitions was found.
Given the much larger number of models studied here might allow to find even deeper relations between
seemingly different fiber types.


\section*{Acknowledgments}

We would like to thank Philip Candelas, I\~naki Garc\'ia-Etxebarria,
and Tom Pugh for interesting discussions.
The work of T.G.~and J.K.~is supported by a
grant of the Max Planck Society. V.B.~is supported by the EPSRC grant
BKRWDM00.

\appendix

\section{List of Non-Toric Non-Abelian Gauge Groups} 
\label{a:groups}

In this appendix we list the non-toric non-Abelian gauge groups that are present if the
coefficients $a_i$ defining the complete intersection are chosen generically. In order to determine
these singularities we computed the Weierstrass forms of the genus-one curves and factorized
$f$, $g$ and $\Delta$. The vanishing degrees along an irreducible factor then determine
the singularity over the vanishing locus of that factor. We quote \autoref{t:kodaira}
from \cite{Grassi:2011hq} for a dictionary to translate the vanishing degrees into the Kodaira type.
Since the total number of singularities we find is very large, we have split up our
results into tables \ref{t:etype}, \ref{t:dtype}, \ref{t:atype1} and \ref{t:atype2}.
Note that we do not include the disconnected spaces corresponding to direct product
nef partitions.

\begin{table}[h!]
{\footnotesize
\begin{center}
\begin{tabular}{|c|c|c|c|c|c|} \hline
&$\operatorname{ord}_\Sigma(f)$&$\operatorname{ord}_\Sigma(g)$
&$\operatorname{ord}_\Sigma(\Delta)$
&Eqn.\ of monodromy cover&$\mathfrak{g}(\Sigma)$\\ \hline\hline
$I_2$&$0$&$0$&$2$&
-- & $\mathfrak{su}(2)$ \\ \hline
$I_m$, $m\ge3$&$0$&$0$&$m$&$\psi^2+(9g/2f)|_{z=0}$
& $\mathfrak{sp}(\left[{\frac m2}\right])$ or $\mathfrak{su}(m)$\\ \hline
$I_0^*$&$\ge2$&$\ge3$&$6$&
$\psi^3+(f/z^2)|_{z=0}\cdot\psi+(g/z^3)|_{z=0}$
& $\mathfrak{g}_2$ or $\mathfrak{so}(7)$ or $\mathfrak{so}(8)$
\\ \hline
$I_{2n-5}^*$, $n\ge3$&$2$&$3$&$2n+1$&$\psi^2+\frac14(\Delta/z^{2n+1})(2zf/9g)^3|_{z=0}$
& $\mathfrak{so}(4n{-}3)$ or $\mathfrak{so}(4n{-}2)$ \\ \hline
$I_{2n-4}^*$, $n\ge3$&$2$&$3$&$2n+2$&$\psi^2+(\Delta/z^{2n+2})(2zf/9g)^2|_{z=0}$
& $\mathfrak{so}(4n{-}1)$ or $\mathfrak{so}(4n)$ \\ \hline
$IV^*$&$\ge3$&$  4$  &$  8$& $\psi^2-(g/z^4)|_{z=0}$
& $\mathfrak{f}_4$ or $\mathfrak{e}_6$ \\ \hline
$III^*$&$  3 $&$   \ge5 $&$   9 $& --
& $\mathfrak{e}_7$ \\ \hline
$II^*$&$ \ge4$&$   5   $&$   10 $& --
& $\mathfrak{e}_8$ \\ \hline
\end{tabular}
\end{center}
\smallskip
\caption[Kodaira-Tate classification of singular fibers.]{Kodaira--Tate classification of singular fibers, monodromy covers,
and gauge algebras, taken from \cite{Grassi:2011hq}. The column with the gauge algebras is to be understood as follows: Assume that the defining
equation of the monodromy cover splits into $n$ irreducible pieces. Then the resulting gauge algebra is the $n^{\textrm{th}}$ algebra
listed in the last column.}\label{t:kodaira}
}
\end{table}

\begin{table}[h!]
 \centering
 \begin{tabular}{cc}
 Generic non-toric Kodaira singularities & Occurences \\
 \hline
 \hline
No singularity & $ 88 $ \\
$ IV^* $ & $ 3 $ \\
$ IV^* \times I_2 $ & $ 8 $ \\
$ IV^* \times I_2 \times I_3 $ & $ 9 $ \\
$ IV^* \times I_2^2 $ & $ 4 $ \\
$ IV^* \times I_2^2 \times I_3 $ & $ 4 $ \\
$ IV^* \times I_2^3 \times I_3 $ & $ 1 $ \\
$ IV^* \times I_3^3 $ & $ 1 $ \\
$ IV^* \times I_3^4 $ & $ 1 $ \\
$ III^* \times I_2 $ & $ 2 $ \\
$ III^* \times I_2 \times I_3 $ & $ 4 $ \\
$ III^* \times I_2^2 $ & $ 1 $ \\
$ III^* \times I_2^2 \times I_4 $ & $ 1 $ \\
$ III^* \times I_2^3 \times I_4 $ & $ 1 $ \\
$ II^* \times I_2 \times I_3 $ & $ 1 $ \\
 \end{tabular}
 \caption{List of generic non-toric $E$- and $F_4$-type Kodaira singularities and the
 number of times they occur.}
 \label{t:etype}
\end{table}

\begin{table}[h!]
 \centering
 \begin{tabular}{cc}
 Generic non-toric Kodaira singularities & Occurences \\
 \hline
 \hline
$ I_0^* $ & $ 39 $ \\
$ I_0^* \times I_2 $ & $ 47 $ \\
$ I_0^* \times I_2 \times I_3 $ & $ 15 $ \\
$ I_0^* \times I_2 \times I_3^2 $ & $ 4 $ \\
$ I_0^* \times I_2^2 $ & $ 27 $ \\
$ I_0^* \times I_2^2 \times I_3 $ & $ 17 $ \\
$ I_0^* \times I_2^2 \times I_4 $ & $ 5 $ \\
$ I_0^* \times I_2^2 \times I_4^2 $ & $ 4 $ \\
$ I_0^* \times I_2^3 $ & $ 15 $ \\
$ I_0^* \times I_2^3 \times I_4 $ & $ 4 $ \\
$ I_0^* \times I_2^4 $ & $ 2 $ \\
$ I_0^* \times I_2^4 \times I_4 $ & $ 3 $ \\
$ I_0^* \times I_2^5 $ & $ 2 $ \\
$ I_1^* $ & $ 9 $ \\
$ I_1^* \times I_2 $ & $ 20 $ \\
$ I_1^* \times I_2 \times I_3 $ & $ 9 $ \\
$ I_1^* \times I_2^2 $ & $ 13 $ \\
$ I_1^* \times I_2^2 \times I_3 $ & $ 8 $ \\
$ I_1^* \times I_2^2 \times I_3^2 $ & $ 2 $ \\
$ I_1^* \times I_2^3 $ & $ 4 $ \\
$ I_1^* \times I_2^3 \times I_3 $ & $ 2 $ \\
$ I_2^* \times I_2 $ & $ 3 $ \\
$ I_2^* \times I_2 \times I_3 $ & $ 7 $ \\
$ I_2^* \times I_2^2 $ & $ 5 $ \\
$ I_2^* \times I_2^2 \times I_4 $ & $ 2 $ \\
$ I_2^* \times I_2^3 \times I_4 $ & $ 2 $ \\
$ I_2^* \times I_2^4 $ & $ 1 $ \\
$ I_2^* \times I_2^5 $ & $ 1 $ \\
$ I_3^* \times I_2 \times I_3 $ & $ 2 $ \\
$ I_3^* \times I_2^2 \times I_3 $ & $ 1 $ \\
$ I_4^* \times I_2^2 \times I_4 $ & $ 1 $ \\
  \end{tabular}
  \caption{List of generic non-toric $G_2$ and $SO$-type Kodaira singularities and the
 number of times they occur.}
 \label{t:dtype}
\end{table}

\begin{table}[h!]
  \centering
  \begin{tabular}{cc}
   Generic non-toric Kodaira singularities & Occurences \\
 \hline
 \hline
$ I_2 $ & $ 263 $ \\
$ I_2 \times I_3 $ & $ 141 $ \\
$ I_2 \times I_3 \times I_4 $ & $ 41 $ \\
$ I_2 \times I_3 \times I_5 $ & $ 12 $ \\
$ I_2 \times I_3 \times I_6 $ & $ 32 $ \\
$ I_2 \times I_3 \times I_7 $ & $ 6 $ \\
$ I_2 \times I_3^2 $ & $ 41 $ \\
$ I_2 \times I_3^2 \times I_4 $ & $ 15 $ \\
$ I_2 \times I_3^3 $ & $ 13 $ \\
$ I_2 \times I_4 $ & $ 136 $ \\
$ I_2 \times I_4^2 $ & $ 4 $ \\
$ I_2 \times I_4^4 $ & $ 1 $ \\
$ I_2 \times I_5 $ & $ 26 $ \\
$ I_2 \times I_6 $ & $ 6 $ \\
$ I_2^2 $ & $ 326 $ \\
$ I_2^2 \times I_3 $ & $ 170 $ \\
$ I_2^2 \times I_3 \times I_4 $ & $ 69 $ \\
$ I_2^2 \times I_3 \times I_5 $ & $ 14 $ \\
$ I_2^2 \times I_3 \times I_6 $ & $ 12 $ \\
$ I_2^2 \times I_3 \times I_7 $ & $ 4 $ \\
$ I_2^2 \times I_3 \times I_8 $ & $ 2 $ \\
$ I_2^2 \times I_3^2 $ & $ 54 $ \\
$ I_2^2 \times I_3^2 \times I_4 $ & $ 15 $ \\
$ I_2^2 \times I_3^2 \times I_5 $ & $ 6 $ \\
$ I_2^2 \times I_3^3 $ & $ 3 $ \\
$ I_2^2 \times I_3^3 \times I_4 $ & $ 2 $ \\
$ I_2^2 \times I_4 $ & $ 134 $ \\
$ I_2^2 \times I_4 \times I_6 $ & $ 6 $ \\
$ I_2^2 \times I_4 \times I_8 $ & $ 8 $ \\
$ I_2^2 \times I_4^2 $ & $ 27 $ \\
$ I_2^2 \times I_4^3 $ & $ 12 $ \\
$ I_2^2 \times I_4^4 $ & $ 1 $ \\
$ I_2^2 \times I_5 $ & $ 28 $ \\
$ I_2^2 \times I_6 $ & $ 22 $ \\
$ I_2^2 \times I_7 $ & $ 2 $ \\
  \end{tabular}
  \caption{List of generic non-toric $Sp$ and $SU$-type Kodaira singularities and the
 number of times they occur, part I.}
 \label{t:atype1}
\end{table}

\begin{table}[h!]
  \centering
  \begin{tabular}{cc}
   Generic non-toric Kodaira singularities & Occurences \\
 \hline
 \hline
$ I_2^3 $ & $ 260 $ \\
$ I_2^3 \times I_3 $ & $ 121 $ \\
$ I_2^3 \times I_3 \times I_4 $ & $ 24 $ \\
$ I_2^3 \times I_3 \times I_5 $ & $ 4 $ \\
$ I_2^3 \times I_3 \times I_6 $ & $ 4 $ \\
$ I_2^3 \times I_3^2 $ & $ 16 $ \\
$ I_2^3 \times I_4 $ & $ 85 $ \\
$ I_2^3 \times I_4 \times I_6 $ & $ 6 $ \\
$ I_2^3 \times I_4^2 $ & $ 10 $ \\
$ I_2^3 \times I_5 $ & $ 10 $ \\
$ I_2^4 $ & $ 133 $ \\
$ I_2^4 \times I_3 $ & $ 30 $ \\
$ I_2^4 \times I_3 \times I_4 $ & $ 2 $ \\
$ I_2^4 \times I_3^2 $ & $ 4 $ \\
$ I_2^4 \times I_4 $ & $ 29 $ \\
$ I_2^4 \times I_4^2 $ & $ 10 $ \\
$ I_2^4 \times I_5 $ & $ 2 $ \\
$ I_2^4 \times I_6 $ & $ 4 $ \\
$ I_2^4 \times I_8 $ & $ 2 $ \\
$ I_2^5 $ & $ 32 $ \\
$ I_2^5 \times I_4 $ & $ 22 $ \\
$ I_2^5 \times I_6 $ & $ 4 $ \\
$ I_2^6 $ & $ 14 $ \\
$ I_2^6 \times I_4 $ & $ 2 $ \\
$ I_2^7 $ & $ 1 $ \\
$ I_2^8 $ & $ 1 $ \\
$ I_3 $ & $ 93 $ \\
$ I_3^2 $ & $ 2 $ \\
$ I_3^3 $ & $ 4 $ \\
$ I_3^3 \times I_6 $ & $ 4 $ \\
$ I_3^3 \times I_9 $ & $ 2 $ \\
$ I_3^4 $ & $ 6 $ \\
$ I_3^4 \times I_6 $ & $ 4 $ \\
$ I_3^5 $ & $ 2 $ \\
$ I_4 $ & $ 95 $ \\
$ I_4^4 $ & $ 1 $ \\
$ I_5 $ & $ 12 $ \\
$ I_6 $ & $ 2 $ \\
 \end{tabular}
  \caption{List of generic non-toric $Sp$ and $SU$-type Kodaira singularities and the
 number of times they occur, part II.}
 \label{t:atype2}
\end{table}
\clearpage


\bibliography{refs}
\bibliographystyle{JHEP}

\end{document}